\documentstyle[multicol,aps,epsf]{revtex}
\begin{document}
\draft
\title{Low Energy Excitations and Phase Transitions in the Frustrated 
Two-Dimensional {\it XY} Model}
\author{Colin Denniston$^{1,2,}$\cite{present} and Chao Tang$^2$}
\address{$^{1}$Department of Physics, Princeton University, 
Princeton, New Jersey 08544}
\address{$^{2}$NEC Research Institute, 4 Independence Way, Princeton,
New Jersey 08540}

\date{\today}
\maketitle

\begin{abstract}
We study the critical properties of the two-dimensional (2D) $XY$ model
in a 
transverse magnetic field with filling factors $f=1/3$ and $2/5$. 
To obtain a comparison with recent experiments, we
 investigate the effect of weak quenched bond disorder for $f=2/5$.  A 
finite-size scaling analysis of extensive Monte Carlo simulations
strongly
suggests that the critical exponents of the phase transition for $f=1/3$
and 
for $f=2/5$ with disorder are those of the pure 2D Ising model.
The relevant low energy excitations are domain walls, and we show that
their properties determine the nature of the phase transition. 
\end{abstract}

\begin{multicols}{2}

\section{Introduction}
In this paper we examine the frustrated $XY$ model in two dimensions for two
different values of the magnetic field representative of 
``commensurate states''.  Experimental realizations of this 
model in the form of two-dimensional arrays of Josephson junctions and 
superconducting wire networks \cite{jja,us1,Sean} can and have been 
constructed and one of the objectives of this work is to understand the
results of these experiments.  A perpendicular magnetic field induces a finite
density of circulating supercurrents, or vortices, within the array.  The
interplay of two length scales -- the mean separation of vortices and the
period of the underlying physical array -- gives rise to a wide variety
of interesting physical phenomena.  Many of these effects show up as
variations in the properties of the finite-temperature superconducting
phase transitions at different fields.  In recent experiments on 
superconducting arrays the critical exponents of a number of these phase 
transitions have been measured \cite{Sean}, opening 
the opportunity to do careful comparison of theory and experiment.   While we
will discuss the model within the context of superconducting networks, the 
model is also closely related to the physics of adsorbed films on substrates 
which impose a periodic potential which differs from the preferred period of 
the adsorbed film. In this work we examine the ground state properties,
low energy excitations, and critical properties of the 2D $XY$ model in the 
densely frustrated regime ($f\gg 0$) for two particular values of the
magnetic field.  In addition, we investigate the effect
of disorder on the ground state and critical properties.  This paper elaborates
and expands upon our previous results reported in Ref.~\cite{us2}.

The Hamiltonian of the frustrated $XY$ model is
\begin{equation}
{\cal H} = - \sum_{\langle ij \rangle} J_{ij} 
\cos(\theta_i-\theta_j-A_{ij}),
\label{ham}
\end{equation}
where $\theta_j$ is the phase on site $j$ of a square $L \times L$ lattice and
$A_{ij}=(2\pi/\phi_0)\int^j_i {\bf A} \cdot d{\bf l}$ is the integral of the
vector potential from site $i$ to site $j$ with $\phi_0$ being the flux 
quantum. The directed sum of the $A_{ij}$ around an elementary plaquette 
$\sum A_{ij}=2\pi f$ where $f$, measured in the units of $\phi_0$, is the
magnetic flux penetrating each plaquette due to the uniformly applied
field.  We focus here on the cases $f=p/q$ with $p/q$ $=$ $1/3$ and $2/5$.

The ground state fluxoid pattern for these $f$ is
shown in Figure~\ref{stairflux}(a) \cite{TeitJay,Halsey}.  The pattern consists
 of diagonal stripes composed of a single line of vortices for
$f={1\over 3}$ and a double line of vortices for $f={2 \over 5}$.
These diagonal lines of vortices can sit on $q$ sub-lattices and, in
addition, there are $q$ more states with the stripes going along the
opposite diagonal for a total of $2 q$ degenerate states.
A common speculation for commensurate-incommensurate 
transitions and the frustrated $XY$ model is that the transition should
 be in the universality class of the q-state (or 2q-state) Pott's model.  
We find that this is not the case because, as discussed below, domain walls 
between the different states vary considerably in both energetic and entropic 
factors.  

The effect of quenched impurities on phase transitions is an important and
fascinating problem.   The ``Harris criterion'' \cite{Harris} indicates that 
the addition of (bond) randomness to systems which exhibit second-order 
transitions in the clean case with a positive specific-heat exponent $\alpha$ 
changes the numerical values of the critical exponents \cite{2ndOrd}.  It has 
also been shown using phenomenological renormalization-group arguments that 
the addition of bond randomness to systems undergoing first-order transitions 
results in a random-field mechanism at any coexistence region which can 
cause the transition to become continuous \cite{HuiBerker}.  
Aizenman and Wehr \cite{AizenmanWehr} have shown quite rigorously that in 2D 
a quenched random field results, quite generally, in the elimination of 
discontinuities in the order parameter conjugate to the fluctuating field.
Most cases where bond disorder has been studied and observed to change the 
order of the transition are for q-state Potts Models, where for $q=8$ 
Chen et al. \cite{Chenetal}  found through extensive Monte Carlo simulations, 
that the first-order transition of the pure model became second-order with 
the critical exponents being consistent with the universality class of the 
two-dimensional Ising model.  Unlike q-state Pott's models with high q, the
frustrated $XY$ system is more readily compared to experiments such as recent 
experimental measurements of critical exponents in superconducting 
arrays \cite{Sean}.

\section{Staircase States}
\label{hstair}

\begin{figure}
\narrowtext
\centerline{\epsfxsize=3in
\epsffile{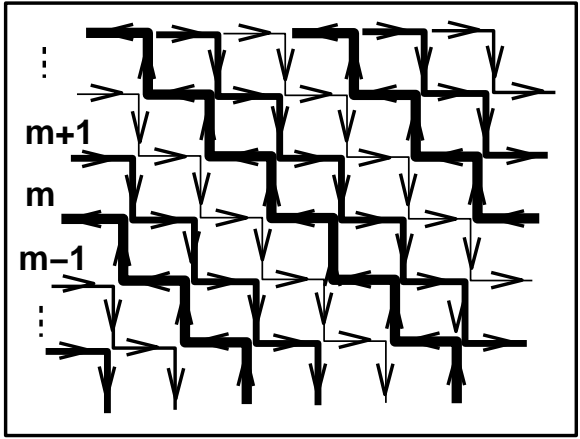}}
\vskip 0.1true cm
\caption{Partition of the square lattice into staircases with the current
flowing up or down the staircases.}
\label{stair}
\end{figure}

The ground states of the Hamiltonian (\ref{ham}) will be among the solutions to
the supercurrent conservation equations 
$\partial {\cal H}/\partial \theta_i=0$:
\begin{equation}
\sum_{j'} \sin(\theta_{j'}-\theta_i-A_{ij'})=0
\label{curr_cons}
\end{equation}
where $j'$ are the nearest neighbors to $i$.  One set of solutions to these
equations was found by Halsey \cite{Halsey} by considering the restriction
to a quasi-one-dimensional case where one has adjoining staircases of current
(see Fig~\ref{stair}(a)).  All gauge invariant phase differences 
$\gamma_m=\theta_{n}-\theta_m-A_{mn}$, within a given staircase are equal and
indexing the staircases by $m$ as shown in Fig~\ref{stair}(a) one finds
\begin{equation}
\gamma_m=\pi f m + \alpha/2-\pi \mathop{\rm nint}[f m+\alpha/(2 \pi)]
\label{stairphase}
\end{equation}
where $\mathop{\rm nint}$ is the nearest integer function, and $\alpha=0$ for 
$f=p/q$ with $q$ odd and $\alpha=\pi/q$ for $q$ even \cite{Halsey}. 

The staircase fluxoid pattern for $f=1/3$ and $2/5$ is 
shown in Fig~\ref{stairflux}(a)\cite{TeitJay,Halsey}.  The pattern consists of 
diagonal stripes composed of a single line of vortices for
$f={1\over 3}$ and a double line of vortices for $f={2 \over 5}$.  
(A vortex is a pla\-quette with unit fluxoid occupation, 
i.e. the phase gains $2 \pi$ when going around the pla\-quette.)  
The stripes shown in Figure~\ref{stairflux} can sit on $q$ sub-lattices, which 
we associate with members of the $Z_q$ group.
They can also go along either diagonal, and we associate these two 
options with members of the $Z_2$ group.  In all, there is a total of $2 q$ 
degenerate states($f=p/q \ne 1/2$).
A common speculation for commensurate-incommensurate 
transitions and the frustrated $XY$ model is that the transition should
 be in the universality class of the q-state (or 2q-state) Pott's model.  
We find that this is not the case because domain walls 
between the different states vary considerably in both energetic and 
entropic factors.

\section{Domain Walls}

\begin{figure}
\narrowtext
\centerline{\epsfxsize=3.2in
\epsffile{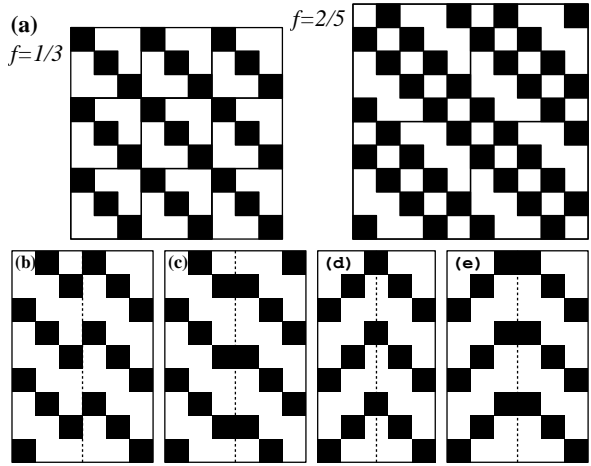}}
\vskip 0.1true cm
\caption{(a) Fluxoid pattern for ground states of $f={1 \over 3}$ and 
$f={2 \over 5}$ (Unit cells are marked by solid lines).  Domain wall
fluxoid pattern for $f={1 \over 3}$: 
(b) shift-by-one wall, (c) shift-by-two wall, (d) herringbone wall, and 
(e) herringbone wall with a shift-by-two (a vortex is shown as a dark square).}
\label{stairflux}
\end{figure}

Figure~\ref{stairflux}(b)-(e) shows the fluxoid pattern for some of the
 domain walls for $f=1/3$.  The domain walls can be classified into two 
types. {\it Shift} walls involve a shift of the vortex pattern across the wall 
(such as in Fig~\ref{stairflux}(b) where the pattern on the right is shifted 
down by one lattice spacings with respect to the pattern on the left) but the 
lines of vortices are still going along the same diagonal.  
{\it Herringbone} walls are walls between states with the vortex stripes going 
along opposite diagonals.  Note that there are $q$ different walls of each 
type.

\begin{figure}
\narrowtext
\centerline{\epsfxsize=3.2in
\epsffile{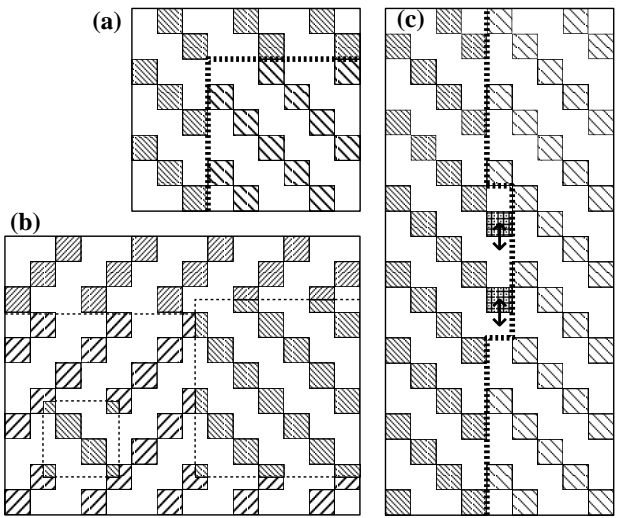}}
\vskip 0.1true cm
\caption{Illustration of several possible bends and kinks in the different
types of domain walls. (a) A 90 degree bend in a $f$ $=$ $1/3$ shift wall 
showing change from shift-by-one to shift-by-two wall.
(b) $f$ $=$ $1/3$ shift-by-one wall branching into two herringbone 
walls.    (c)  Kink in a $f$ $=$ $1/3$ 
shift-by-three wall accomplished by moving the vortex marked in plaid.}
\label{bent_walls}
\end{figure}

These walls also have differing topologies.  A herringbone wall is very 
similar to a domain wall in an Ising model in that it separates two members
of a $Z_2$ group.  It cannot branch into other herringbone walls and a 
90 degree turn in the wall can be accomplished without changing the vortex
pattern, with the caveat that one considers the wall to be composed of sections
of length equal to the distance between the diagonal lines of vortices (see 
Figure~\ref{bent_walls}).  Thus, if one only has herringbone walls in the 
system, the set of possible domain wall configurations is similar to those in 
an Ising model.  Shift walls, on the other hand can branch, both into other 
shift walls (with the constraint that the sum of the shifts on the walls after 
the branch be equal to the original shift) and into a pair of herringbone 
walls, as shown in Figure~\ref{bent_walls}(b).   Shift walls also have an 
associated directionality in the sense that an attempt to make a 90 degree turn
in a shift-by-n wall results in the wall changing to a shift-by-(q-n) wall 
(see Fig.~\ref{bent_walls}(a) for an illustration).  Since different shift 
walls can have quite different energies (see below) one finds that bends such
as the one shown in Fig.~\ref{bent_walls}(a) are energetically highly 
unfavorable as it can change a wall with low energy into a wall with a very
high energy cost.  A more energetically favorable kink in a shift wall can be
formed by displacing a mismatched vortex on the wall in a direction parallel
to the wall\cite{Faloetal}.  This displaces a section of the wall one unit cell
 in the direction perpendicular to the wall (see Figure \ref{bent_walls}(c)).  
Typically, one finds only kinks like these of size one or two lattice 
constants.  Larger kinks start to produce long range distortions in the 
phase field and have higher energy. 

In order to calculate the energies of different structures, we solved the 
equations (\ref{curr_cons}) numerically, using a quasi Newton method, on 
lattices with up to $2.3\times 10^5$ sites with constraints fixing the fluxoid 
occupation of each plaquette (see Appendix \ref{ap:copt}).  
Table~\ref{wall_energies} lists the energy per unit length $\sigma$ for 
straight domain walls between the various ground states at zero temperature
for $f=1/3$ and $2/5$.  One can see from the table that there is 
typically one or two walls with considerably lower energy than any of the 
others.  Some of the patterns of energies seen in the table can be understood
by counting the number of extra vortices in next or next-next-nearest
neighbor plaquettes for the vortices along the wall.  For instance, the 
energy of a $f=1/3$ shift-by-one wall is about twice that of the standard
herringbone wall.  Looking at Figure~\ref{stairflux}(b)-(e) one can see that if
you count the number of next-next-nearest neighbor vortices for vortices along
each side of the wall, the shift-by-one wall has twice as many as the 
herringbone.  Similarly, walls which place vortices on nearest neighbor sites 
tend to be of a higher energy, or may not even be stable.  While this does give
a rough guide to the pattern of energies, it does not allow a strong comparison
 of walls with differently spaced vortices.  

One can see from Figure~\ref{bent_walls}(b) that a shift wall can be viewed as 
two adjacent, or {\it bound} herringbone walls. For $f=1/3$ the energy 
of two herring\-bone walls is less than that of a single shift wall and hence, 
the shift walls should be unstable to breaking up into herringbone walls. 
As a result, one expects that in the $f=1/3$ case if the temperature is high
enough for domain walls to enter the system, the herringbone 
walls should be the only walls present at large length scales. 
While for $f=1/3$, herringbone walls are the only stable walls, this is
 not true for $f=2/5$.  For $f=2/5$ it is energetically favorable
 for two herringbone walls to bind and form a shift-by-one or shift-by-three 
wall.  This can lead to more complex domain wall structures and has an 
important impact on the nature of the finite temperature phase transition.  
These issues will be addressed in more detail below.

\begin{figure}
\narrowtext
\centerline{\epsfxsize=3in
\epsffile{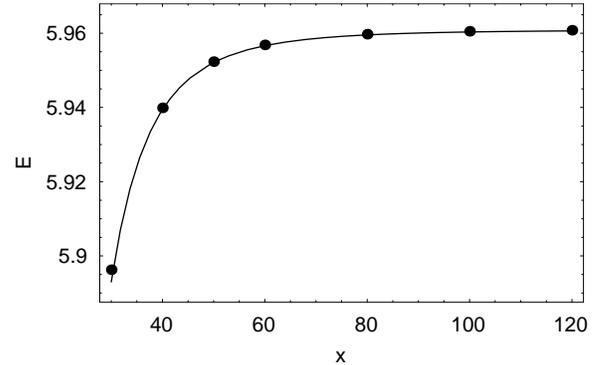}}
\vskip 0.1true cm
\caption{Energy of a square domain of size $15\times 15$ in a system with
periodic boundary conditions of size $x \times x$ for $f=2/5$.  The
line is the fit $5.961081(7)$ $-1.086(1) (15/x)^4$.}
\label{dom_dom2}
\end{figure}

We also numerically calculated the energy of domain walls that are not 
straight.  Figure \ref{dom_dom2} shows the energy of a square closed domains, 
formed from herringbone walls, of linear dimension $L$ unit cells in a system 
of size 120x120 with periodic boundary conditions.  We see that to a very 
good approximation, the energy scales linearly in $L$.  One can, however,
work out some corrections to this linear dependence due to the change in the
vortex density at the corner of the domains.  For instance in 
Fig.~\ref{bent_walls}(b) the vortices at opposite corners of the square domain
have either an extra next-nearest neighbor vortex or a missing next-nearest
neighbor.  From a distance, this gives a quadrapole moment to the domain.  As
the $3\times 3$ domain shown in Fig~\ref{bent_walls}(b) is the basic building
block of larger domains, one can conclude that larger differently shaped 
domains will not have a lower moment (i.e. they will be neutral and have no 
dipole moment).  The interaction of two such quadrapole domains at a distance 
$x$, large compared to it's size $L$ goes like $-6 J_{eff} (L/x)^4$ (if one 
assumes an isotropic (which is not really true) interaction of two ``corner'' 
charges like $J_{eff} \ln x$).  In addition, the self energy of a square 
quadrapole goes like $-2 J_{eff} \ln \sqrt{2}+2 J_{eff}\ln L$.  

\begin{figure}
\narrowtext
\centerline{\epsfxsize=3in
\epsffile{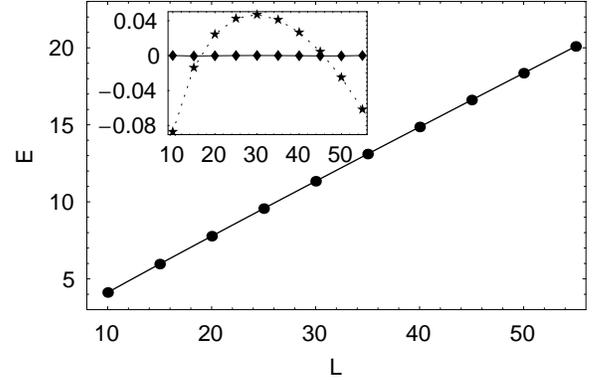}}
\vskip 0.1true cm
\caption{Energy of a square domain of size $L\times L$ in a system
with periodic boundary conditions of size $120\times 120$ for $f=2/5$.  The 
line is the fit 
$-0.0268(25)$ $+0.344797(68) L$ $+0.301(1) \ln L$ $-1.28(3)(L/120)^4$. 
The inset shows the residuals for a linear fit (stars) and the fit
including the quadrapole corrections (diamonds).}
\label{dom_dom1}
\end{figure}

Figure \ref{dom_dom1} shows the interaction of some square domains.  One sees 
that the quadrapole correction is measurable and fits the expected functional 
form quite well, but that the constants $J_{eff}$ do not match what one would 
expect from an isotropic calculation.  In fact, the system is not really
equivalent to an isotropic 2D Coulomb gas, in that the direction along the 
diagonal lines of vortices in the staircase state is not equivalent to the
direction perpendicular to the vortex lines.  We have also calculated the 
energies of rectangular domains and some other less regular shapes and they 
have qualitatively similar (same functional form) behavior.

The next question is whether or not the quadrapole interaction is likely to
be relevant.  One can use an argument similar to that used to argue for a 
transition in the unfrustrated $XY$ model.  If you consider the interaction
free energy contribution of the quadrapole interaction, it should contain 
the energetic part $-A/r^4$ and an entropic part $-B T \ln (\pi r^2)$ from 
confining the quadrapoles to have a separation less than $r$ (One could do
a more accurate calculation of the entropy but it will still have a $\ln r$
dependence).  At the distances at which the $-A/r^4$ form is valid, the $\ln r$
term wins all the time (at finite $T$) and hence one can argue that the
quadrapole interactions should not be relevant.

\section{Spin Waves}

At low enough temperature, domains should be small, and one is tempted to 
expand the energy about the ground state configuration.  In this treatment,
the periodic character of the angles is neglected, but the existence of 
long range order in the vortex lattice partly justifies this method.  The
model is replaced by a so-called spin wave approximation which involves
expanding the Hamiltonian to 2nd order in $\Delta_{ij}$, where 
$\theta_{ij}=\theta_{ij}^{(0)}+\Delta_{ij}$ and $\theta_{ij}^{(0)}$ is a 
ground state configuration:
\begin{eqnarray}
{\cal H}\approx{\cal H}^{(0)}
&+&\sum_{ij} \left({\partial {\cal H} \over \partial \theta_{ij}} \right)^{(0)} \Delta_{ij}\nonumber\\
&+& {1\over 2}\sum_{ij}\sum_{kl}\Delta_{kl}\left({\partial^2 {\cal H} \over \partial \theta_{kl}\partial \theta_{ij}} \right)^{(0)}\Delta_{ij}
\label{swham}
\end{eqnarray}
By definition, $(\partial {\cal H}/ \partial \theta_{ij})^{(0)}=0$ and we just
have a quadratic form.  The free energy per site associated with (\ref{swham})
is
\begin{eqnarray}
{\cal F}&=&-{1 \over \beta} \ln {\cal Z}_{sw}\nonumber\\
&=& -{1 \over \beta}\ln \left[ \int \prod_i d\Delta_{\bf x} \right. \nonumber\\
& &  \qquad\qquad\times\left.\exp\left(-{\beta\over 2}\sum_{\bf x,x'}\Delta_{\bf x}\left({\partial^2 {\cal H} \over \partial \theta_{\bf x}\partial \theta_{\bf x'}} \right)^{(0)}\Delta_{\bf x'}\right)\right]\nonumber\\
&=&-{1 \over \beta}\ln \left(det {{\bf J}\over 2 \pi}\right)^{-1/2}
\end{eqnarray}
where ${\bf J}$ is the Jacobian matrix,
${\bf J_{x,x'}}=\partial^2 {\cal H}/ \partial \theta_{\bf x}\partial \theta_{\bf x'}$.
The spin wave correlation function is 
\begin{eqnarray}
G_{sw}(&{\bf x_1},&{\bf x_2}) \nonumber\\
 &=&\langle \exp[i(\Delta_{\bf x_1}-\Delta_{\bf x_2})] \rangle\\
&=&{\cal Z}_{sw}^{-1} \int \prod \Delta_{\bf x} \nonumber\\
& &\times\exp\left[-{\beta\over 2}\sum_{\bf x,x'}\Delta_{\bf x}{\bf J_{x,x'}} \Delta_{\bf x'}+
i(\Delta_{\bf x_1}-\Delta_{\bf x_2})\right]\nonumber\\
&=&\exp\left(-{1\over 2\beta}X({\bf x_1,x_2})^T {\bf J}^{-1} X({\bf x_1,x_2})\right),\nonumber
\end{eqnarray}
where $X({\bf x_1,x_2})$ is a vector with $+1$ and $-1$ in positions 
${\bf x_1}$ and ${\bf x_2}$ respectively and zeros everywhere else.

For the unfrustrated case \cite{KTrefs} ${\bf J}$ is just the discretization of
 the Laplacian operator (i.e. ${\bf J} \phi\approx \nabla^2 \phi$, where 
$\nabla^2 \phi=\partial^2 \phi/\partial x^2+\partial^2 \phi/\partial y^2$ and
the partial derivatives are replaced with a finite-difference formula
$\partial^2 \phi/\partial x^2 \approx (\phi(x_{i+1},y_i)-2\phi(x_i,y_i)+\phi(x_{i-1},y_i))/a^2$, and $a$ is the lattice constant).  As a result,  
${1\over 2}X({\bf x,x'})^T {\bf J}^{-1} X({\bf x,x'})$ 
can be approximated by the Green's function for the Poisson equation, 
$$g(r)={1 \over 2 \pi} \ln {r \over r_0},$$
where $r_0=a/(2 \sqrt{2} e^\gamma)$, $a$ is the lattice spacing and
$\gamma=0.577216\cdots$ is Euler's constant.  
This yields 
\begin{eqnarray}
G_{sw}({\bf x_1,x_2})
& \sim & \exp(-1/2 \pi \beta \ln (|{\bf x_1}-{\bf x_2}|/r_0) \nonumber\\
& = & (r_0/|{\bf x_1}-{\bf x_2}|)^{1/2\pi \beta}
\end{eqnarray}
So the correlation function of the spin wave fluctuations decreases according
to a power law behavior.  This algebraic decay of the spin wave correlation
function is broken by the unbinding of vortex-antivortex pairs at the 
Kosterlitz-Thouless transition \cite{KTrefs}.  

\begin{figure}
\narrowtext
\centerline{\epsfxsize=3.in
\epsffile{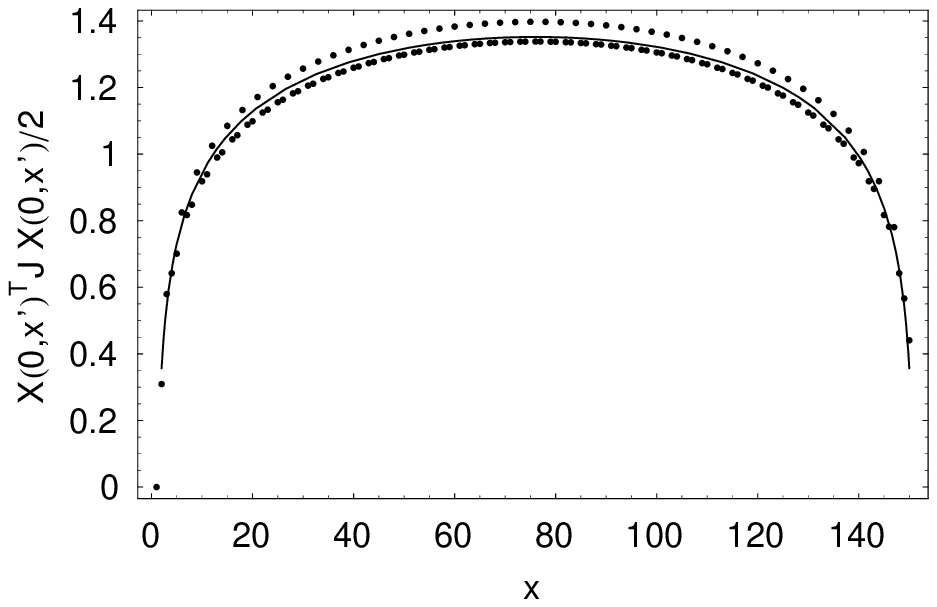}}
\vskip 0.1true cm
\caption{${1\over 2}X({\bf x,x'})^T {\bf J}^{-1} X({\bf x,x'})$, the 
lattice ``Green's'' function,
for $f$ $=$ $1/3$ along a slice in the $x$-direction in a finite size system 
with periodic boundary conditions along the direction of the slice.  The 
line indicates a fit to $A(\ln x+\ln(L-x))$.}
\label{swcor}
\end{figure}

In the general frustrated case, ${\bf J}$ is not a discrete Laplacian.  The 
question is, do we get something similar?  The $1/r^4$ interaction of the 
domains studied in the previous section suggests that we do.
Figure \ref{swcor} shows 
${1\over 2}X({\bf x,x'})^T {\bf J}^{-1} X({\bf x,x'})$ 
for $f=1/3$ along a slice in the $x$-direction in a finite size system 
with periodic boundary conditions along the direction of the slice. 
The envelope of this curve is well described by the sum of two logarithmic
functions, $\ln x+\ln (L-x)$ (where the second term comes from the periodic
boundary conditions).  In addition to this logarithmic part, there is a
periodic oscillation, coinciding with the underlying vortex lattice.  In 
addition to this obvious oscillation, the phase of the oscillation depends on
the initial ${\bf x}$ (The correlation function is not just a function of
$({\bf x}-{\bf x'})$).  The effect of this initial ${\bf x}$ dependent phase
at long distances should not be important.  However, distortions centered
on nearby sites, and between rows of vortices can partially cancel due to
this phase difference.  There is also an anisotropy between the directions
perpendicular and parallel to the diagonal lines of vortices.  This 
anisotropy can, however, be removed in a continuum picture by rescaling the
coordinates.

This modified lattice ``Green's'' function also has an impact on vortex 
interactions.  The presence of the logarithmic part ensures that the 
overall flux balancing ($f=\langle n_i \rangle$ where $n_i$ is the vortex
occupation of plaquette $i$) is maintained.  However the vortex interaction
energy should contain an oscillating component coinciding with the underlying
vortex lattice.  The effect of such a component
is not entirely clear, especially as the amplitude of the oscillation does 
not decay away at large distances.  Conventional wisdom would suggest that as 
long as we still have the logarithmic interaction of vortices, they should 
still undergo a Kosterlitz-Thouless type of unbinding transition and an 
associated jump in the helicity modulus \cite{KTrefs}.  As we shall see in the 
next section, it is not entirely clear whether or not this actually happens.  
It might be interesting to try to go through and derive the Kosterlitz 
recursion relations with the oscillations as some sort of perturbation to 
see if it is relevant, although it seems unlikely to do anything but 
renormalize the core energies.

\section{$\lowercase{f}=1/3$}

The fluxoid pattern for the two lowest energy walls at $f$ $=$ ${1\over 3}$  
was shown in Figure~\ref{stairflux}(b) and (d).  One can see from 
Figure \ref{bent_walls}(b) 
that a shift wall can be viewed as two adjacent, or {\it bound} herringbone 
walls. For $f={1 \over 3}$ the energy of two herring\-bone walls is less than 
that of a single shift wall and hence, the shift walls are unstable and break 
up into herringbone walls. As a result, we confine our discussion of the 
$f={1\over 3}$ case to the herringbone walls as other walls should not be 
present at large length scales.  The energy cost for dividing an $L \times L$ 
lattice into two domains separated by a solid-on-solid (SOS) wall stretching 
from one side of the system to the other is
\begin{equation}
{\cal H}_{single} \{z\}= b \sigma L+b \sigma \sum_k |z_k-z_{k-1}|.
\label{singleSOS}
\end{equation}
The height variables $z_k$ take on integer values ($b=3$ is the
shortest length segment).  The partition
 function, ${\cal Z}=\sum_{\{z_k\}}\exp (-{\cal H}/T)$ can be evaluated 
either by the transfer matrix method or recursively (see 
Appendix \ref{ap:SOS})\cite{Forgacsetal}.  The interfacial free energy per column is
$
{\cal F}=T \ln [ e^{b \sigma/T}\tanh(b \sigma/(2T))].
$
The zero crossing of ${\cal F}$ gives an estimate of the critical 
temperature. Plugging in the values for the $f={1 \over 3}$ herringbone wall 
gives $T_c=0.19 J$, in remarkable agreement with the value $T_c=0.22 J$ 
found in the Monte Carlo simulations described below. 
 
Being similar to Ising walls, herringbone walls cannot branch into other 
herringbone walls, thus the set of possible domain wall configurations is 
similar to those in an Ising model.  We label the fraction of the system in
state $(s,j)$ as $m_{s,j}$, where $s=\pm1$ denotes the member of $Z_2$, and 
$j=1,2,3$ denotes the member of $Z_3$.  Below the transition, one state 
$(s,i)$ spans the system.  On this state sit fluctuating domains, bounded by 
herringbone walls, of each of the states $(-s,1), (-s,2),$ and $(-s,3)$ in 
equal numbers; so the $Z_3$ symmetry is broken for the $(s,j)$ states, but not
 for the $(-s,j)$ states.  As the transition is approached from below, the 
domains occupied by the $(-s,j)$ states grow, with smaller domains 
 of the $(s,j)$ states within them.  At the transition,
the $Z_2$ symmetry between the $\pm s$ states is restored and, as a result,
the $Z_3$ symmetry for the $(s,j)$ states is also restored.  

The Monte Carlo simulations used a heat bath algorithm with system sizes of 
$20 \le L \le 96$.  We computed between $10^7$ and $3\times 10^7$ 
Monte Carlo steps (complete lattice updates) with most of the data taken
 close to $T_c$.  Data from different temperatures was 
combined and analyzed using histogram 
techniques \cite{FerrenburgSwend}(see Appendix \ref{ap:mc}).

If the largest fraction of the system is in state $(s,i)$, then we
have three Ising order parameters, 
$M_j=(m_{s,i}-m_{-s,j})/(m_{s,i}+m_{-s,j}), \, j=1\cdots 3.$ 
On average, these $M_j$ are the same so we just take the average as $M$.  To 
calculate the $m_{\sigma,i}$, we examine the 
Fourier transform of the vortex density $\rho_{k\pm}$ at the 
reciprocal lattice vectors ${\bf k_\pm}={\pi \over 3}(1,\pm1)$ of the ground 
state vortex lattices.  Starting from the definition of the Fourier transform,
 and using the vortex states given above, one finds 
\begin{equation}
{\rho_{k\pm}\over\rho_g}=m_{\pm 1,1}+m_{\pm 1,2}e^{i 2 \pi/3}+
	m_{\pm 1,3}e^{-i 2 \pi/3},
\end{equation} 
where $\rho_g$ is the modulus in the ground state.  In practice, $\rho_{k\pm}$ 
is reduced by small short-lived regions 
which don't quite match any of the six states.  Since this effect is the
same for all states, it cancels when calculating $M$. 
Using the real and imaginary parts of $\rho_{k\pm}$ in addition to 
$\sum_j m_{\pm1,j}$, calculated from the direct vortex lattice as in 
\cite{LeeLee}, we can find the five independent $m_{\sigma,j}$.

In addition to the energy and order parameter, several other quantities were 
calculated from the Monte Carlo data using the corresponding 
fluctuation-dissipation relations:
\begin{eqnarray}
{C \over k_B} &=& 
  {K^2 \over L^2} (\langle E^2 \rangle - \langle E \rangle ^2),\nonumber\\
\chi &=& K L^2 (\langle M^2 \rangle -\langle M \rangle ^2), \nonumber\\
{\partial \ln \langle M^n \rangle \over \partial K} &=& 
{\langle M^n E \rangle \over \langle M^n \rangle}-\langle E \rangle,
\label{flucdis}
\end{eqnarray}
where $K=J/k_B T$. In addition to the discrete order 
parameter, we also followed the helicity modulus defined by
$Y_{x,y}=\partial^2 {\cal F}/\partial \phi^2 |_\phi=0$, where ${\cal F}$ is 
the free energy density and $\phi$ is a twist in the boundary condition along 
the $x$ or $y$ direction.  The helicity modulus also follows a 
fluctuation-dissipation relation which is used in calculating it from the data:
\begin{eqnarray}
Y_x &=& {1\over L^2} \left\langle \sum_{\langle {\bf r},{\bf r'} \rangle} [({\bf r}-{\bf r'})\cdot \hat{\bf x}]^2 \cos(\theta_{\bf r}-\theta_{\bf r'}-A_{{\bf r},{\bf r'}}) \right\rangle\nonumber\\
&-& {\beta J^2 \over L^2} \left\langle \left[ \sum_{\langle {\bf r},{\bf r'} \rangle} [({\bf r}-{\bf r'})\cdot \hat{\bf x}] \sin(\theta_{\bf r}-\theta_{\bf r'}-A_{{\bf r},{\bf r'}}) \right]^2 \right\rangle \nonumber\\
&+& {\beta J^2 \over L^2} \left\langle \sum_{\langle {\bf r},{\bf r'} \rangle} [({\bf r}-{\bf r'})\cdot \hat{\bf x}] \sin(\theta_{\bf r}-\theta_{\bf r'}-A_{{\bf r},{\bf r'}}) \right\rangle^2\nonumber\\
& &
\end{eqnarray}
where $\langle {\bf r},{\bf r'} \rangle$ denotes nearest neighbor pairs.

To determine the critical exponents for the transition we make use of finite
size scaling \cite{FSS}. Following standard 
arguments, one assumes that for a second-order transition, the singular part 
of the free energy, $F(t,h)$, near the transition is dominated by a term that 
changes under a change of scale according to the ansatz
$$
F(t,h)=\lambda F(\lambda^s t, \lambda^r h)
$$
where $t=(T-T_c)/T_c$ and $h$ is an applied field which couples to the order
parameter $M$ (so $h$ is not the true magnetic field here).  From this, one
can derive the scaling form of the order parameter, specific heat, 
susceptibility, etc. using the standard relations, $M=-\partial F/\partial h$,
 $C=-T \partial^2 F/\partial t^2$, $\chi=\partial m/\partial h$, etc.  
If one takes the special case $h=0, \lambda=|t|^{-1/s}$ one can relate $r,s$ 
to the standard exponents $\alpha$ for the specific heat, $\beta$ for the
order parameter, and $\gamma$ for the susceptibility as $s=1/(\alpha-2)$,
$r=(\gamma+\beta)/(\alpha-2)$ and $\alpha+2 \beta+\gamma=2$.  If one takes
the case $h=0$ and $\lambda=L^{(\alpha-2)/\nu}$, where $\nu$ is the exponent
for the divergence of the correlation length, one obtains the relations
for finite size scaling:
\begin{eqnarray}
M^n &=& L^{-\beta/\nu} {\cal M}_n(x_t),\nonumber\\
C &=& L^{\alpha/\nu} {\cal C}(x_t),\nonumber\\
\chi &=& L^{\gamma/\nu} {\cal X}(x_t).
\label{scalingI}
\end{eqnarray}  
where $x_t=tL^{1/\nu}$ is the temperature scaling variable.  Using relations 
\ref{scalingI} one can also derive \cite{Chenetal,FerrLan}
\begin{eqnarray}
{\partial \langle M \rangle \over \partial T} &=&
	L^{{1-\beta \over \nu}} {\cal D}(x_t), \nonumber\\
{\partial \ln\langle M \rangle \over \partial T} &=&
	L^{1/\nu} {\cal Q}(x_t).
\label{scalingII}
\end{eqnarray}
For a finite lattice the peak in, for example the specific heat, scales with 
system size like $C_{max} \propto L^{\alpha/\nu}$ and occurs at the 
temperature where the scaling function ${\cal C}(x_t)$ is maximum so that
$$
{{\cal C}(x_t) \over dx_t}|_{x_t=x_t^*}=0.
$$
This defines the finite-lattice transition temperature $T_c(L)$ by the 
condition $x_t=x_t^*$ so that $T_c(L)=T_c+T_cx_t^*L^{-1/\nu}$.  In general the
finite-lattice transition temperature calculated from different quantities 
differs slightly but extrapolates to the same $T_c$ in the limit of large $L$.

\begin{figure}
\narrowtext
\centerline{\epsfxsize=3in
\epsffile{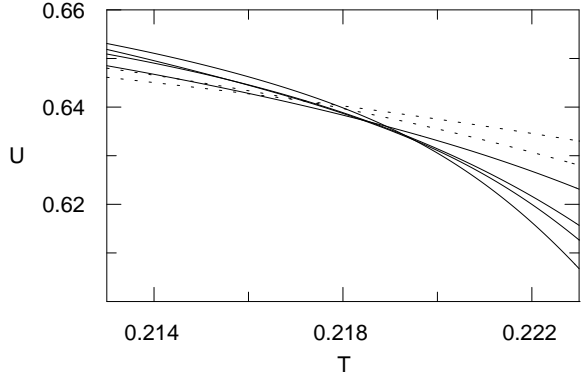}}
\vskip 0.1true cm
\caption{$f=1/3$ Binder's cumulant $U$ vs $T$ for $L=36$ to $L=84$ (smaller
 $L$ shown as dotted lines).}
\label{Bind}
\end{figure}

A very accurate way of locating the transition temperature is by using 
Binder's cumulant \cite{Binder},
$$ 
U=1-\langle M^4 \rangle/(3 \langle M^2 \rangle^2),
$$
shown in Figure~\ref{Bind}.  For system sizes large enough to
 obey finite-size scaling, this quantity is size independent at the critical 
point.  From Fig.~\ref{Bind} we find $T_c=0.2185(6)J$.  $T_c$ can also be 
determined from the scaling equation for the temperature at the peak of 
thermodynamic derivatives such as the susceptibility, 
$T_c(L)=T_c+a L^{-1/\nu}$. We find these other methods give $T_c$ in agreement
 with that from $U$.

\begin{figure}
\narrowtext
\centerline{\epsfxsize=3in
\epsffile{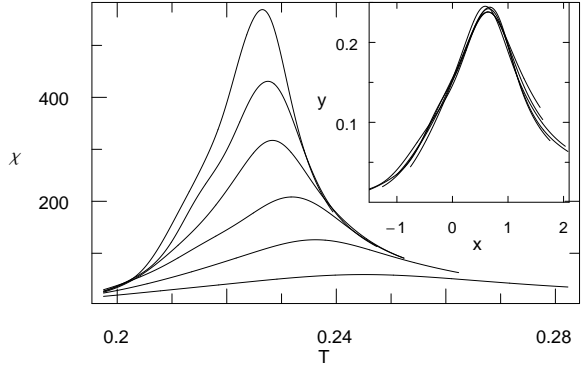}}
\vskip 0.1true cm
\caption{$f=1/3$ $\chi$ vs $T$ for $L=36$ to $L=84$ and
scaling collapse of this data (inset) where $x=(T-T_c)L^{1/\nu}$, 
$y=\chi L^{-\gamma/\nu}$, $\nu=1$, and $\gamma={7 \over 4}$.}
\label{Chi}
\end{figure}

\begin{figure}
\narrowtext
\centerline{\epsfxsize=3in
\epsffile{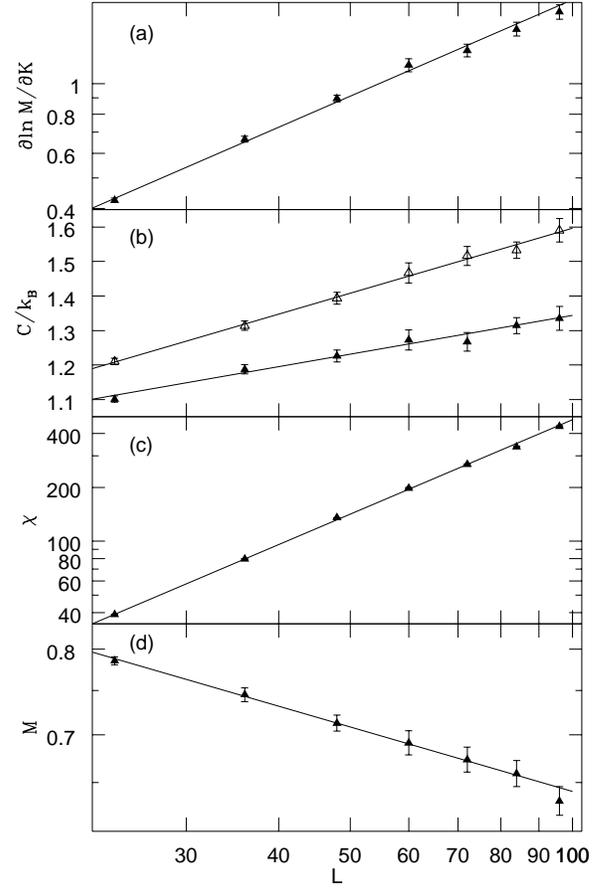}}
\vskip 0.1true cm
\caption{Finite size scaling plots for $f={1 \over 3}$.
(a) logarithmic derivative of $M$  at $T_c$ vs $L$, 
(b) specific heat maximum (hollow) and at $T_c$ (solid) vs $L$,
(c) $\chi$ at $T_c$ vs $L$, and
(d) $M$ at $T_c$ vs $L$.}
\label{f13scale}
\end{figure}

Finite size scaling \cite{FSS} at $T_c$ applied to 
$\partial \ln M / \partial K$ gives $1/\nu=1.011\pm 0.029$, and to
 the susceptibility $\chi$ gives $\gamma/\nu=1.758 \pm 0.013$, and to $M$ gives
 $\beta/\nu=0.14\pm 0.02$.  These exponents are determined from the slopes of 
the lines shown in Fig.~\ref{f13scale} which plots the values of these 
quantities at the critical point as function of $L$.  These exponents are in 
excellent agreement with the Ising values $\nu=1$, $\gamma={7 \over 4}$, and 
$\beta={1 \over 8}$.  Fig.~\ref{Chi} shows the collapse of the raw data onto 
the scaling function (inset) for $\chi$.

\begin{figure}
\narrowtext
\centerline{\epsfxsize=3in
\epsffile{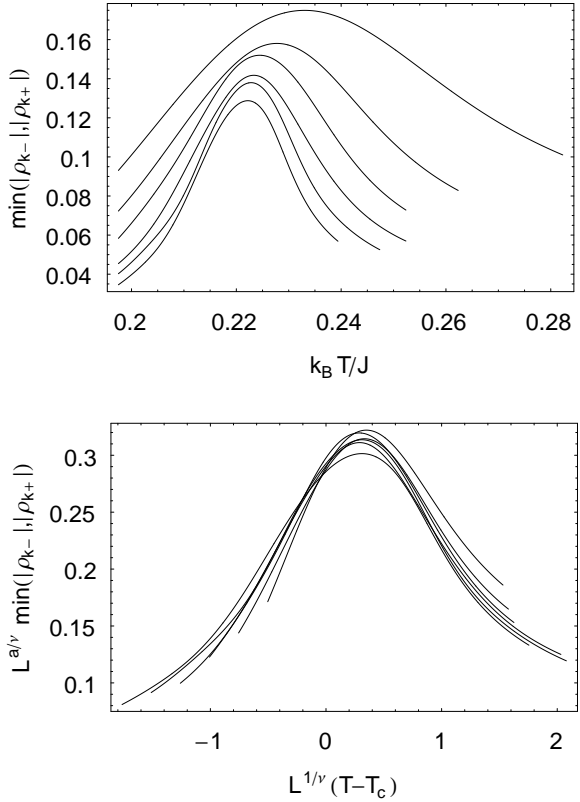}}
\vskip 0.1true cm
\caption{left:min$(\rho_{k+},\rho_{k-})$ versus $k_B T/J$.  Note that data
from larger $L$ are smaller: min$(\rho_{k+},\rho_{k-})$ vanishes as 
$L\rightarrow \infty$ as indicated by the finite-size scaling plot (right) 
which shows a reasonable collapse for 
min$(\rho_{k+},\rho_{k-})\sim L^{a/\nu}$ with $a/\nu=-0.20\pm0.02$.}
\label{rhokmin}
\end{figure}

\begin{figure}
\narrowtext
\centerline{\epsfxsize=2.9in
\epsffile{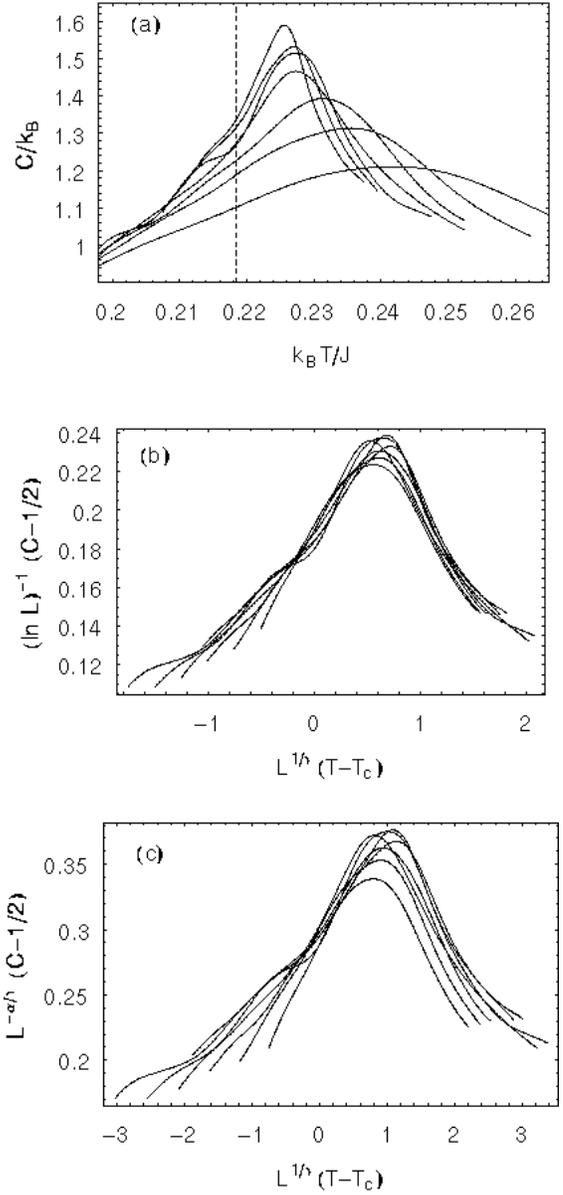}}
\vskip 0.1true cm
\caption{(a) Specific heat for $L=24$ to $L=96$.  The dashed line 
indicates $T_c$.  Note that the shoulder which appears for intermediate 
lattice sizes goes away for the two largest $L$.  This makes the scaling
of $C$ not as good as for the other variables. 
(b) Scaling collapse of data shown in (a). 
(c) Power law scaling found by Lee and Lee for smaller system sizes, applied to
 the data shown in (a).  The logarithmic scaling shown in (b) gives a better 
collapse of the data.  In particular the lower curve in (c), corresponding to
the scaled $L=96$ data is separating from the pack.}
\label{Cf13}
\end{figure}

Two previous examinations of the $f={1 \over 3}$ case \cite{Faloetal,Grest} 
suggested a continuous transition but did not measure critical exponents.
Lee and Lee \cite{LeeLee} claim to find separate, closely spaced transitions, 
for the breaking of $Z_2$ and $Z_3$.  One explanation for their conflicting 
results comes from the small system sizes ($L \le 42$) used in their analysis.
  Below the transition, if the dominant state is $(s,i)$, in small systems 
you often do not see all three of the $(-s,j)$ states in the system at the 
same time.  Figure \ref{rhokmin} illustrates this effect.  The minimum of
$(\rho_{k+},\rho_{k-})$ is a measure of the $Z_3$ symmetry breaking for the 
$(-s,j)$ states and this goes to zero as $L\rightarrow \infty$.  
The finite value of min$(\rho_{k+},\rho_{k-})$ for small $L$ can give the 
impression of separate transitions for small systems (If a
measured parameter contains a contribution from min$(\rho_{k+},\rho_{k-})$ it's
derivatives can have a double peaked structure from the derivative of 
min$(\rho_{k+},\rho_{k-})$).  One must take care in the choice of order 
parameter to ensure that this contribution is not biasing the results.  For
example we found that the derivative of the Ising order parameter used 
in \cite{LeeLee}, $M'=m_1+m_2+m_3-m_4-m_5-m_6$ has a double peaked structure 
for intermediate lattice sizes that does not completely go away until $L=96$.  
This makes $M'$ an unsuitable choice of order parameter for finite-size 
scaling.  This is also the probable cause of the 
presence of a shoulder in the specific heat at intermediate system sizes 
\cite{LeeLee}.  For larger $L$, we see this shoulder merge with 
the main peak and for $L=84$ and $96$ it is no longer clearly discernible
(see Fig.\ref{Cf13}).

\begin{figure}
\narrowtext
\centerline{\epsfxsize=2.9in
\epsffile{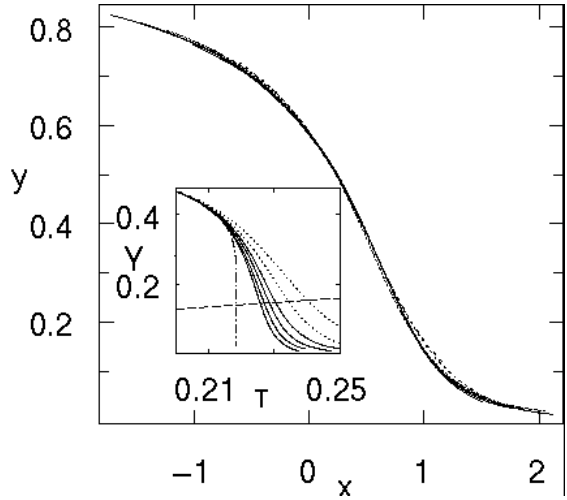}}
\vskip 0.1true cm
\caption{Scaling collapse of $Y$ where
$x=(T-T_c)L^{1/\nu}$, $y=Y L^{\beta/\nu}$, $\nu=1$, and $\beta={1 \over 8}$.
Inset: raw data (solid and dotted), ${2 \over \pi} T$ (dashed), and 
$a|T-T_c|^\beta$ (dot-dashed).}
\label{Ycollapse}
\end{figure}

The helicity modulus $Y$ is the quantity most closely related to experimental 
measurements\cite{KTrefs}.  For $f \ne 0$, the scaling of the $I$-$V$ curves 
found in experiments is consistent with domain wall activation 
processes \cite{Sean}.
The theory of Nelson and Kosterlitz for the $f=0$ case predicts that $Y$ 
should come down in a characteristic square-root cusp and
then jump with a universal value, $2k_BT_{KT}/\pi$.  However, we find an
exceptionally good fit (Fig~\ref{Ycollapse}) of our data to 
$Y-Y_0$ $=$ $L^{-\beta/\nu}$ ${\cal M}((T-T_c)L^{1/\nu})$ with $\nu=1$, 
$\beta$ $=$ ${1 \over 8}$, and $Y_0$ $=$ $0$, which is the scaling form of 
$M$.  Clearly, $Y$ is affected strongly by fluctuations in $M$ and attempting
 to fit scaling relations for the $f=0$ case 
\cite{LeeLee} without taking this into account seems questionable.  We see two
 possible interpretations of our result.  The first is that $Y$ only 
receives contributions from the ordered part of the lattice. 
So comparisons with the $f$ $=$ $0$ case should examine $Y_m=Y/M$.  
$Y_m\approx 0.58$ at the transition implying a larger than 
universal jump.  Alternatively, one can say that although $Y$ is brought 
down by fluctuations in $M$, it should still jump when it
crosses the universal value, $2k_BT/\pi$.  Extrapolating 
the observed behavior of $Y$ gives 
$Y_{L\rightarrow \infty}$ $=$ $a|T-T_c|^\beta$. 
This crosses the value of the universal jump at $T_{KT}-T_c \approx 10^{-6}$.
Although we do not see evidence for a jump, a 
difference in transition temperatures of $10^{-6}$ would not 
lead to any observable effects for the system sizes studied here.

\section{$\lowercase{f}=2/5$}

While for $f={1\over 3}$, herringbone walls are the only stable walls, this is
 not true for $f={2\over 5}$.  For 
$f={2\over 5}$ it is energetically favorable for two herringbone walls to bind
 and form a shift-by-one or shift-by-three wall.  Binding does, however, 
have an entropic cost.  To see if these walls are bound we consider the 
following model for two SOS walls:
\begin{eqnarray}
{\cal H}_{d} \{\Delta,z\}&=& 
	\sum_k \{(2b\sigma+u_\parallel\delta_{z_k,0})
	+b\sigma |z_k-z_{k-1}|\nonumber\\
& & +(2b\sigma+u_\perp \delta_{z_k,0}) \Delta_k
	+V_r(\{\Delta,z\})\}.
\label{doubleSOS}
\end{eqnarray}
$z_k$ is the separation of the walls ($z_k \geq 0$),
$\Delta_k$ is the number of vertical steps the two walls take in the same
direction in the k'th column ($-\infty < \Delta_k <\infty$).  $u_\parallel$
and $u_\perp$ are the binding energies parallel and perpendicular
to the wall.   At this stage we take $V_r=0$.  The solution to such a model
is discussed in Appendix \ref{ap:SOS}.  A ground state 
eigenvector $\psi_\mu(z)$ $=$ $e^{-\mu z}$, where $1/\mu$ is the localization 
length, or typical distance separating the lines, characterizes the bound 
state of the two lines.  $\mu=0$ defines the unbinding transition at $T_b$.  
For the cases of interest, one finds $T_b=0.398 J$ for the shift-by-one walls 
and $T_b=0.442 J$ for the shift-by-three walls.  In addition, the free energy 
for these walls crosses zero before they unbind.  Hence, at the transition, 
defined by the point at which the walls enter the system, we expect a 
branching domain wall structure similar to the $q\ge 5$ Pott's models where a 
first order phase transition occurs.  Technically, this is a mean field 
argument for the interfaces but, since the interfaces are extended objects it
should give a reasonable picture of the order of $T_b$ for the interfaces
and $T_c$.

\begin{figure}
\narrowtext
\centerline{\epsfxsize=3in
\epsffile{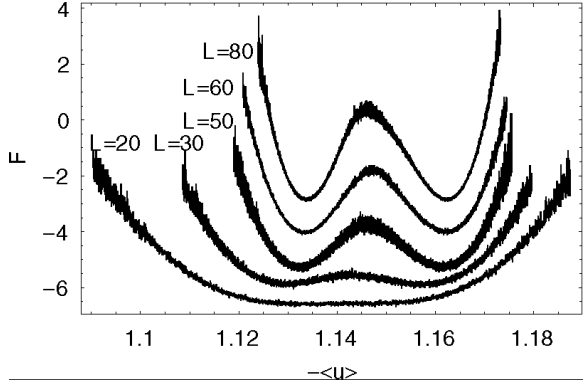}}
\vskip 0.1true cm
\caption{Free energy as function of the negative of the energy per site
for $f=2/5$ ($\delta=0$).  A constant has been added to the curves in order
to separate them.}
\label{d0f25barr}
\end{figure}

In their Monte Carlo simulations, Li and Teitel \cite{LiTeitel} observed 
hysteresis of the internal energy when the temperature was cycled around the 
transition and used this as an argument for a first order 
transition at $f={2\over 5}$.  The most direct indication of a first order 
transition is the presence of a free energy barrier between the 
ordered and disordered states which diverges as the system size increases 
\cite{LeeKosterlitz}.  The free energy as a function of energy is obtained
 using ${\cal F}_L(E)=-\ln P_L(E)$ where $P_L(E)$ is 
the probability distribution for the energy generated by Monte Carlo 
simulation of a $L\times L$ system. 
Figure~\ref{d0f25barr} shows the growth in this barrier as the 
system size increases from $L=20$ to $80$ giving clear evidence for the 
first order nature of the transition.    

\begin{figure}
\narrowtext
\centerline{\epsfxsize=3in
\epsffile{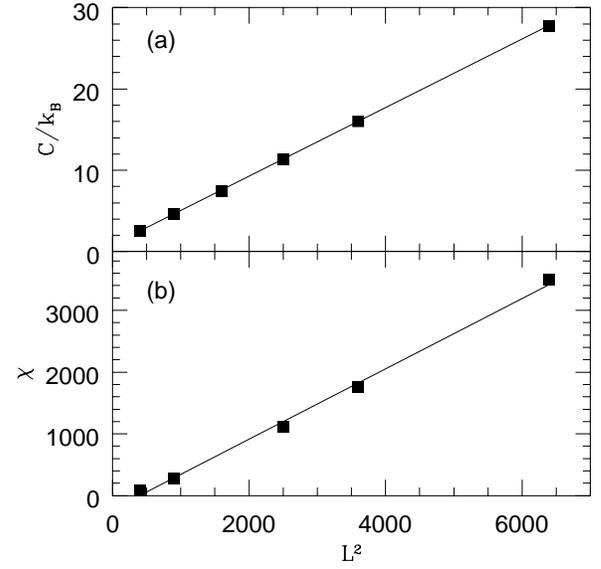}}
\vskip 0.1true cm
\caption{(a) Specific heat verses $L^2$ and (b) susceptibility versus $L^2$.
Errors are comparable to the symbol sizes.}
\label{f25CChi}
\end{figure}

Since there is no diverging characteristic length to which the linear 
dimension $L$ could be compared at a first order transition, one finds that it
 is simply the volume $L^d$ that controls the size effects \cite{1storder}.  
One thus finds 
$$
C_{max},\chi \propto L^d
$$
for a first-order transition.  Figure \ref{f25CChi} shows the specific heat 
as a function of $L^2$ for the $f=2/5$ clean system.  The linear fit 
(solid line) clearly shows the expected first-order scaling behavior.  Similar
behavior can be seen in the susceptibility as shown in the Figure.  From the
positions of the peaks as a function of $L$ we 
obtain $T_c=0.2127(2) J$.

\section{Disorder and the $\lowercase{f}=2/5$ Phase Transition}

We now consider the effects of disorder on the $f={2\over 5}$ phase transition.
 Taking the couplings in the Hamiltonian (\ref{ham}) as 
$J_{ij}=J(1+\epsilon_{ij})$, the $\epsilon_{ij}$ are chosen randomly
from a Gaussian distribution with a standard deviation $\delta$.
Due to variations of the phase differences across the bonds, a specific 
realization of random bonds may favor a certain sub-lattice for the ground 
state, creating an effective random field. To quantify the effect, we placed 
the fluxoid configuration of the ground states down on $10\,000$ 
separate realizations of the disorder and allowed the continuous degrees of 
freedom (the phases) to relax and minimize the energy.
We find that the changes in energy from the $\delta=0$ case fit
a Gaussian distribution with mean $-0.5 \delta^2 L^2$ and standard
deviation $\delta L$.  The difference in energy between states which 
were degenerate in the clean system is the measure of the 
random field.  This difference centers on zero and has
a standard deviation of $0.75 \delta L$ for two states related
by a shift and $0.57 \delta L$ for two states with vortex rows
along opposite diagonals.  The effect of random fields on discrete degrees of 
freedom in 2D is marginal \cite{ImryMa}. For $D>2$ 
there is a critical randomness above which random fields cause the formation of
 domains in the ground state of size $\sim \xi_{\rm rf}$.  Aizenman and Wehr 
have shown that this critical randomness is zero in 2D
\cite{AizenmanWehr}.  Yet, their result does 
not preclude the possibility that 
$\xi_{\rm rf}$ is so large as to be unobservable in a finite sized sample.  
Indeed,
 experiments on superconducting arrays have found apparent phase transitions, 
including scaling behavior \cite{Sean} in sample sizes of order 
$1000 \times 1000$.  In our simulations with disorder at $\delta \leq 0.1$,
 all systems had a low temperature state with the order parameter approaching 
unity.  We will, therefore, ignore the effects of random fields for 
$\delta \leq 0.1$ assuming that $\xi_{\rm rf}$ is larger than the sample size.

At any coexistence point of the clean system, random {\it bonds} result in 
different regions of the system experiencing average couplings slightly above 
or below the critical coupling.  As a result, at any given temperature the 
system will predominantly prefer 
either the ordered or disordered state wiping out the coexistence region and 
leaving only a continuous transition \cite{ImryMa,HuiBerker,AizenmanWehr}.  
It has been conjectured \cite{Chenetal} 
that critical random Potts models are equivalent to Ising models.  Kardar et 
al. \cite{KardarII} suggested a possible mechanism for this effect.  Their 
position space renormalization group approximation suggests that the 
probability of loop formation in the fractal interface of the clean system 
vanishes marginally at a transition dominated by random bonds.  The 
interface may have some finite width due to a froth of bubbles of different
phases, but under renormalization a linear critical interface is obtained and, 
hence, an Ising transition appears.  

The fluxoid configurations from our simulations suggest that for large
enough disorder, ($\delta>\delta_f$) the interface is really linear, not just 
in the renormalized sense.  $\delta_f$ can be estimated by placing a random 
potential $V_r$ in Eq.~\ref{doubleSOS}.  Ignoring the terms involving 
$\Delta_k$, one obtains the model for wetting in the presence of disorder, 
solved by Kardar\cite{Kardar} in the continuum limit.  
He obtained a new length scale due to randomness,
$$
1/\kappa=2T^3 / K \delta^2
$$
where K is the renormalized stiffness related to the interfacial free energy
$\sigma(\theta)$ by $K=\sigma(0)+d^2\sigma(\theta)/d\theta^2 |_0$ where 
$\theta$ is a small tilt angle of the interface.  For an Ising-like interface
$K\approx T/a \sinh[b \sigma/T-\ln\coth(b \sigma/(2T))]$ \cite{Forgacsetal}.
The unbinding transition is lowered and is now defined by the condition 
$\mu-\kappa=0$.  As $T_b$ decreases, it
eventually hits the transition temperature for the first order phase transition
 observed in the clean system.  At this point any branched domain wall 
structure is unstable.  This is just the last step in a process in which the 
effective linear interface becomes narrower as disorder increases.  In the 
vicinity of this ``final'' (mean-field) unbinding, the Ising-type behavior of 
the system should be readily visible at any length scale.

\begin{figure}
\narrowtext
\centerline{\epsfxsize=3in
\epsffile{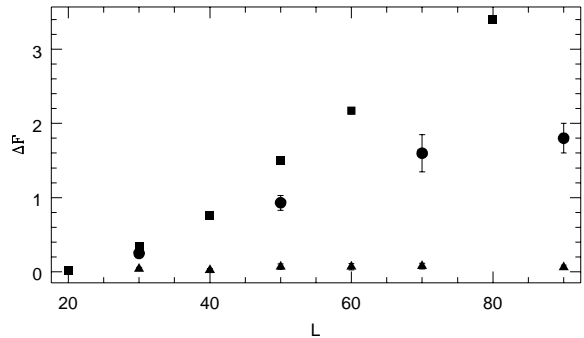}}
\vskip 0.1true cm
\caption{Free energy barrier vs system size for 
$f={2 \over 5}$ and $\delta=0$ (squares), $0.05$ (circles) and 
$0.10$ (triangles).  $\delta$ is the bond disorder strength.}
\label{f25barr}
\end{figure}

We have done a Monte Carlo analysis with bond disorder values of 
$\delta=0.05$ and $0.1$.  Since we are dealing with quenched disorder, we are
interested in averaged quantities; for instance the free energy is 
\begin{equation}
{\cal F}=-k_BT [\ln Z]_{av}
\end{equation}
where the square brackets indicate an average over different realizations of
disorder.  Since most quantities of interest involve derivatives of the 
free energy, to calculate the average value of a thermodynamic 
quantity, we first calculate it for a given realization of the disorder and 
then do a configurational average over 10 to 15 realizations for 
$\delta=0.1$ and seven realizations for $\delta=0.05$. 
Figure~\ref{f25barr} shows the free energy barrier for $f={2\over 5}$ as a 
function of system size in the for $\delta=0.05$, and $0.1$.  
For $\delta=0.05$, the barrier first grows with system size and then levels 
off.  At $\delta=0.1$ the free energy barriers are essentially zero, 
indicating a continuous transition and that the system sizes are large enough 
to apply finite size scaling.  Here, we follow the finite-size scaling methods
used in \cite{Chenetal}.

\begin{figure}
\narrowtext
\centerline{\epsfxsize=3in
\epsffile{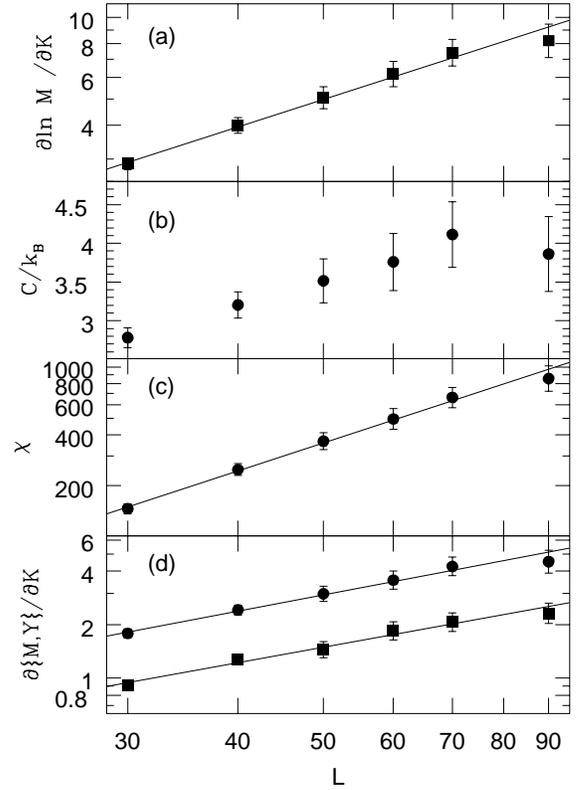}}
\vskip 0.1true cm
\caption{Finite size scaling plots for 
$f={2 \over 5}, \delta=0.1$: (a) logarithmic derivative of $M$ vs
 $L$, (b) $C/k_B$ vs $L$, (c) $\chi$ vs $L$, and (d) $\partial M/\partial K$
and $\partial Y/\partial K$ vs $L$}
\label{scale}
\end{figure}

Figure~\ref{scale} shows the peak values of $\partial \ln M /\partial K$ and 
$\chi$ as a function of $L$. The slopes of these 
plots give $1/\nu=1.05(12)$ and $\gamma/\nu=1.70(12)$.  A similar analysis of
$\partial M/\partial K$ gives $(1-\beta)/\nu=0.94(10)$ \cite{us2}.  As in 
the $f=1/3$ case, the helicity modulus appears to track the order parameter
$M$.  Within errors, these exponents are what one
would expect from an Ising model. Experiments at $f={2\over 5}$ \cite{Sean} 
also found a continuous transition and measured the critical exponents
$\nu=0.9(5)$ and the dynamic critical exponent $z=2.0(5)$, consistent with
an Ising transition.

\section{Conclusions}

In conclusion, we find that the nature and universality class of the phase
transitions are quite sensitive to the proximity of the binding transition for 
the lowest energy domain walls.  For $f=1/3$ the lowest energy walls are never 
bound and the transition is Ising-like.  For $f=2/5$ domain walls can lower 
their free energy by binding to each other, resulting in
 a first order phase transition.  Disorder weakens this binding and 
changes the transition to be continuous and Ising-like.  These results are 
consistent with the continuous phase 
transition and critical exponents observed experimentally for 
$f=2/5$ \cite{Sean}. 

We thank M. Aizenman, P. Chandra, J.M. Kosterlitz, X.S. Ling, and D. Huse and 
for useful discussions.

\appendix
\section{Constrained Optimization for Vortex Lattices}
\label{ap:copt}

Minima of the Hamiltonian (\ref{ham}) satisfy Equations (\ref{curr_cons}).  
However, these equations are written in terms of the $\theta_j$ variables and
the locations of the vortices does not enter explicitly.  This is quite 
inconvenient as one finds that the zero temperature energies of the system
are almost entirely dictated by the vortex structure.  By this we mean that 
given the position of all the vortices, the phases appear to be uniquely 
determined (up to an overall constant) by the minimization conditions.  This
can be made more explicit by working with the gauge invariant phase differences
\begin{eqnarray}
\gamma_{i,j}&=&\theta_{i,j}-\theta_{i,j-1}-
  {2\pi \over \phi_0}\int^{(i,j)}_{(i,j-1)} {\bf A} \cdot d{\bf l},\nonumber\\
\alpha_{i,j}&=&\theta_{i-1,j}-\theta_{i,j}-
  {2\pi \over \phi_0}\int^{(i-1,j)}_{(i,j)} {\bf A} \cdot d{\bf l},
\end{eqnarray}
where $\theta_{i,j}$ is the phase on the site at row $i$ column $j$ of the 
lattice.  This introduces an extra variable per site (instead of just 
$\theta_{i,j}$ now we have $\gamma_{i,j}$ and $\alpha_{i,j}$) and a 
compensating constraint that
\begin{equation}
\gamma_{i,j}-\gamma_{i-1,j}+\alpha_{i,j}-\alpha_{i,j-1}-2 \pi (f-n_{i,j})=0.
\label{gauge_const}
\end{equation}
That is to say, the sum of the gauge invariant phase differences around any
plaquette must equal the magnetic flux through the plaquette $2 \pi f$, plus
an integer multiple $n_{i,j}$ of $2 \pi$.  If the gauge invariant phase 
differences are restricted to a range of $2 \pi$ such as $[-\pi,\pi)$ then
$n_{i,j}$ measures the vortex occupancy of the plaquette and is typically 
$0$ or $\pm 1$ with the sign depending on the sign of $f$. 

One then rewrites Equations (\ref{curr_cons}) in terms of the gauge invariant
phase differences to get
\begin{equation}
\sin \gamma_{i,j}-\sin \gamma_{i,j+1}+\sin \alpha_{i+1,j}-\sin \alpha_{i,j}=0.
\end{equation}
If disorder is added, the random couplings should be included here.
These, in addition to Eq.'s (\ref{gauge_const}) give $2 M N$ equations (for a 
lattice of $M\times N$ unit cells) for the $2 M N$ unknown gauge invariant
phase differences.  The vortex pattern $\{ n_{i,j}\}$ is now an input and 
stays fixed.  When periodic boundary conditions are imposed one finds that
two of these equations are not independent.  Two more convenient conditions
to impose closure are
\begin{eqnarray}
\sum_{j=1}^M \sin \alpha_{N,j}-I_c &=& 0, \nonumber\\
\sum_{i=1}^N \sin \gamma_{i,1}-I_r &=& 0,
\end{eqnarray}
where $I_c$ is the net current flowing down the columns of the lattice and
$I_r$ is the net current flowing along the rows.  In all cases found, the 
lowest energy state corresponded to $I_{r,c}=0$.  

The above equations can now be organized into the form 
${\bf F}(\{\gamma_{i,j},\alpha_{i,j}\})=0$ as 
\begin{eqnarray}
F_1 &=& \sum_{i=1}^N \sin \gamma_{i,1}-I_r, \nonumber\\
F_{2M(i-1)+2j-1} &=& \gamma_{i,j}-\gamma_{i-1,j} \nonumber\\
 & & \qquad +\alpha_{i,j}-
	\alpha_{i,j-1}-2 \pi (f-n_{i,j}), \nonumber\\
F_{2M(i-1)+2j} &=& \sin \gamma_{i,j}-\sin \gamma_{i,j+1} \nonumber\\
 & & \qquad +\sin \alpha_{i+1,j}-\sin \alpha_{i,j}, \nonumber\\
F_{2 M N} &=& \sum_{j=1}^M \sin \alpha_{N,j}-I_c.
\label{equations}
\end{eqnarray}
If we define ${\bf x}$ to have elements $x_{2M(i-1)+2j-1}=\gamma_{i,j}$ and 
$x_{2M(i-1)+2j}=\alpha_{i,j}$ ($i=1\cdots N$ and $j=1\cdots M$) then the 
solution to (\ref{equations}) can be found using Newton's method which involves 
iteratively solving 
\begin{equation}
{\bf J}\cdot \delta {\bf x}=-{\bf F}
\label{Newtmethod}
\end{equation}
 and updating ${\bf x}$,
\begin{equation}
{\bf x}_{new}={\bf x}_{old}+\delta {\bf x},
\end{equation}
where the Jacobian $J_{i,j}=\partial F_i/\partial x_j$.

The set of equations (\ref{Newtmethod}) can be very large (we solved systems 
with up to $2.3\times 10^5$ sites which means Eq.(\ref{Newtmethod}) represents 
about half a million simultaneous equations).  In addition, we need to solve
these systems very fast, especially when disorder is added and averages over
tens of thousands of solutions are needed.  This is made possible by the special
form of the Jacobian matrix:
\begin{equation}
\arraycolsep=2pt
\renewcommand{\arraystretch}{0.25}
{\bf J}=
\left[
\begin{array}{cccccc|cccccc|cccccc|cccccc|cccccc}
\cdot &\cdot & & & & &\cdot & & & & & &\cdot & & & & & &\cdot & & & & & &\cdot & & & & &\\
\cdot &\cdot &\cdot & & & & &\cdot & & & & & & & & & & & & & & & & & &\cdot & & & &\\
&\cdot &\cdot &\cdot & & & & &\cdot & & & & & & & & & & & & & & & & & &\cdot & & &\\
& &\cdot &\cdot &\cdot & & & & &\cdot & & & & & & & & & & & & & & & & & &\cdot & &\\
& & &\cdot &\cdot &\cdot & & & & &\cdot & & & & & & & & & & & & & & & & & &\cdot &\\
\cdot & & & &\cdot &\cdot & & & & & &\cdot & & & & & & & & & & & & & & & & & &\cdot\\ \cline{1-30}
\cdot & & & & & &\cdot &\cdot & & & &\cdot &\cdot & & & & & & & & & & & & & & & & &\\
&\cdot & & & & &\cdot &\cdot &\cdot & & & & &\cdot & & & & & & & & & & & & & & & & \\
& &\cdot & & & & &\cdot &\cdot &\cdot & & & & &\cdot & & & & & & & & & & & & & & \\
& & &\cdot & & & & &\cdot &\cdot &\cdot & & & & &\cdot & & & & & & & & & & & & & &\\
& & & &\cdot & & & & &\cdot &\cdot &\cdot & & & & &\cdot & & & & & & & & & & & & &\\
& & & & &\cdot &\cdot & & & &\cdot &\cdot & & & & & &\cdot & & & & & & & & & & & & \\ \cline{1-30}
& & & & & &\cdot & & & & & &\cdot &\cdot & & & &\cdot &\cdot & & & & & & & & & & &\\
& & & & & & &\cdot & & & & &\cdot &\cdot &\cdot & & & & &\cdot & & & & & & & & & &\\
& & & & & & & &\cdot & & & & &\cdot &\cdot &\cdot & & & & &\cdot & & & & & & & & &\\
& & & & & & & & &\cdot & & & & &\cdot &\cdot &\cdot & & & & &\cdot & & & & & & & &\\
& & & & & & & & & &\cdot & & & & &\cdot &\cdot &\cdot & & & & &\cdot & & & & & & &\\
& & & & & & & & & & &\cdot &\cdot & & & &\cdot &\cdot & & & & & &\cdot & & & & & &\\ \cline{1-30}
& & & & & & & & & & & &\cdot & & & & & &\cdot &\cdot & & & &\cdot &\cdot & & & & &\\
& & & & & & & & & & & & &\cdot & & & & &\cdot &\cdot &\cdot & & & & &\cdot & & & &\\
& & & & & & & & & & & & & &\cdot & & & & &\cdot &\cdot &\cdot & & & & &\cdot & & &\\
& & & & & & & & & & & & & & &\cdot & & & & &\cdot &\cdot &\cdot & & & & &\cdot & &\\
& & & & & & & & & & & & & & & &\cdot & & & & &\cdot &\cdot &\cdot & & & & &\cdot &\\
& & & & & & & & & & & & & & & & &\cdot &\cdot & & & &\cdot &\cdot & & & & & &\cdot\\ \cline{1-30}
\cdot & & & & & & & & & & & & & & & & & &\cdot & & & & & &\cdot &\cdot & & & &\cdot \\
&\cdot & & & & & & & & & & & & & & & & & &\cdot & & & & &\cdot &\cdot &\cdot & & &\\
& &\cdot & & & & & & & & & & & & & & & & & &\cdot & & & & &\cdot &\cdot &\cdot & &\\
& & &\cdot & & & & & & & & & & & & & & & & & &\cdot & & & & &\cdot &\cdot &\cdot &\\
& & & &\cdot & & & & & & & & & & & & & & & & & &\cdot & & & & &\cdot &\cdot &\cdot\\
& & & & & & & & & & & & & & & & & & & & & & & & &\cdot & &\cdot & &\cdot\\
\end{array}
\right]
\end{equation}
where the dots represent the non-zero elements.  We see that ${\bf J}$ is very
nearly band diagonal.  In fact ${\bf J}$ can be written as
\begin{equation}
{\bf J}={\bf A}+{\bf U}\cdot {\bf V}^T
\end{equation}
where ${\bf A}$ is the band diagonal part of ${\bf J}$ (the same three matrix
diagonal blocks as ${\bf J}$) and ${\bf U}$ and ${\bf V}$ are $N\times 2M$ 
matrices (as opposed to $2MN\times 2MN$).  I should point out here that the 
method described below has a speed that is proportional to $NM^2$ so that the
axes of the lattice should always be chosen so that $M \leq N$ for efficient
operation.  ${\bf U}$ and ${\bf V}$ have the form
\begin{eqnarray}
{\bf U}^T&=&  \left[
\arraycolsep=1.72pt
\renewcommand{\arraystretch}{0.25}
\begin{array}{cccccc|cccccc|cccccc|cccccc|cccccc}
{}_1 & & & & & &\; & & & & & &\; & & & & & &\; & & & & & &\cdot & & & & &\\
& {}_0 & & & & & &\; & & & & & &\; & & & & & &\; & & & & & &\cdot & & & &\\
& & {}_1 & & & & & &\; & & & & & &\; & & & & & &\; & & & & & &\cdot & & &\\
& & & {}_0 & & & & & &\; & & & & & &\; & & & & & &\; & & & & & &\cdot & &\\
& & & & {}_1 & & & & & &\; & & & & & &\; & & & & & &\; & & & & & &\cdot &\\
& & & & & {}_0 & & & & & &\; & & & & & &\; & & & & & &\; & & & & & &\cdot\\
\end{array}
\right],\nonumber\\
{\bf V}^T&=& \left[
\arraycolsep=1.72pt
\renewcommand{\arraystretch}{0.25}
\begin{array}{cccccc|cccccc|cccccc|cccccc|cccccc}
{}_0 & & & & & &\; & & & & & &\cdot & & & & & &\cdot & & & & & &\cdot & & & & &\\
& {}_1 & & & & & &\; & & & & & &\; & & & & & &\; & & & & & &\cdot & & & &\\
& & {}_0 & & & & & &\; & & & & & &\; & & & & & &\; & & & & & &\cdot & & &\\
& & & {}_1 & & & & & &\; & & & & & &\; & & & & & &\; & & & & & &\cdot & &\\
& & & & {}_0 & & & & & &\; & & & & & &\; & & & & & &\; & & & & & &\cdot &\\
& & & & & {}_1 & & & & & &\; & & & & & &\; & & & & & &\; & & & & & &\cdot\\
\end{array}
\right].\nonumber\\
\end{eqnarray}
The first two blocks of ${\bf U}$ and ${\bf V}^T$ have the nonzero elements 
indicated and and the remaining blocks of ${\bf U}$ are from the first block 
column of ${\bf J}$ and the remaining blocks of ${\bf V}$ are from the first 
block row of ${\bf J}$.

The solution of a band diagonal system ${\bf A}\cdot{\bf x}={\bf b}$ is 
considerably simpler than solving a general linear system of $2 M N$ equations.
Not only that, but the $LU$ factorization of ${\bf A}$ has the same storage
requirements as ${\bf A}$ which can be stored in a packed storage scheme holding
only the central nonzero band.  In order to solve our slightly more general
problem we make use of the {\it Woodbury formula} \cite{NumRec}:
\begin{eqnarray}
{\bf J}^{-1}&=&({\bf A}+{\bf U}\cdot {\bf V}^T)^{-1}\nonumber\\
&=&{\bf A}^{-1} \nonumber\\
& &-\left[{\bf A}^{-1}\cdot{\bf U}\cdot({\bf 1}+{\bf V}^T\cdot{\bf A}^{-1}\cdot{\bf U})^{-1}\cdot{\bf V}^T\cdot{\bf A}^{-1}\right].\nonumber\\
\label{woodbury}
\end{eqnarray}
Since storage of ${\bf A}^{-1}$ is not practical (the inverse does not preserve
the band structure of the matrix), we must make use of (\ref{woodbury}) in the
following way, as described in \cite{NumRec}:
To solve the linear equation
\begin{equation}
({\bf A}+{\bf U}\cdot {\bf V}^T)\cdot \delta {\bf x}=-{\bf F}
\end{equation}
first solve the $2M+1$ auxiliary problems
\begin{equation}
{\bf A}\cdot {\bf Z}={\bf U},
\end{equation}
and
\begin{equation}
{\bf A}\cdot {\bf y}=-{\bf F}.
\end{equation}
This can be done by $LU$ factorizing ${\bf A}$ {\it once} and then using the
factorization to solve all the systems simultaneously.  Routines from
LAPACK \cite{netlib} can make this very fast and efficient.  Next, do the 
$2M\times 2M$ matrix inversion
\begin{equation}
{\bf H}\equiv({\bf 1}+{\bf V}^T \cdot {\bf Z})^{-1}.
\end{equation}
In terms of these quantities, the solution is given by
\begin{equation}
\delta{\bf x}={\bf y}-
	{\bf Z}\cdot\left[{\bf H}\cdot({\bf V}^T\cdot{\bf y})\right].
\end{equation}

In order to start Newton's method, one needs a good initial guess.  This is
provided by patching together the staircase state solutions described in 
section \ref{hstair}.  In addition, care must be taken to ensure that 
the gauge invariant phase differences do not wander out of $[-\pi,\pi)$.  There
are a number of options one can use if a phase difference wanders out of range.
One is to just pin the solution at $\pm \pi$.  This is not a great solution as
this is not really a minima of the unconstrained Hamiltonian.  Another solution
is to just add or subtract $2 \pi$ and continue iterating Newton's method.  This
can cause a jump in the errors on one of the equations which may result in a 
large change in ${\bf x}$ at the next step which may or may not be beneficial.  
Another solution is to replace the phase difference with the value on the other
branch of the $\arcsin$ function on $[-\pi,\pi)$.  This causes no change in the
error on the current conservation equations and produces a smaller change in the
corresponding Eq.(\ref{gauge_const}).  Many of these problems can often be 
avoided by taking a step in the Newton direction but with smaller length, 
especially in the initial stages, using a dynamic step length algorithm similar
to those described in \cite{NumRec}.

\section{Solid on Solid Models}
\label{ap:SOS}
\begin{figure}
\centerline{\epsfxsize=3.1in
\epsffile{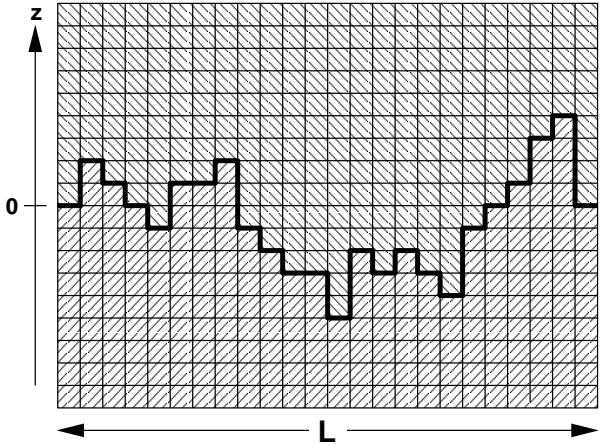}}
\vskip 0.1true cm
\caption{Solid on solid interface.  Overhangs and bubbles are ignored in the 
SOS model and interface configurations can be described in terms of 
integer-valued height variables measured from the straight, $T=0$ configuration
of the interface.}
\label{one_sos}
\end{figure}

A good review of interface models is given in \cite{Forgacsetal}.  
Here we briefly discuss the cases relevant to our situation.  The SOS model
of an interface ignores overhangs and bubbles and configurations can be 
described in terms of integer-valued height variables whose values are 
measured from the $T=0$ position of the interface (see Figure~\ref{one_sos}).  
The energy cost for dividing an $L \times L$ lattice into two domains 
separated by a solid-on-solid (SOS) wall stretching from one side of the 
system to the other is
\begin{equation}
{\cal H}_{single} \{z\}= b \sigma L+b \sigma \sum_k |z_k-z_{k-1}|.
\label{simpleSOS}
\end{equation}
The height variables $z_k$ take on integer values ($b$ is the shortest length 
segment).  The partition function, ${\cal Z}=\sum_{\{z_k\}}\exp (-{\cal H}/T)$ 
can be easily evaluated by change of variables, $\Delta_i=z_i-z_{i-1}$ so that
$$
{\cal Z}=\prod_{k=1}^L e^{-\beta b \sigma} 
	\sum_{\Delta_k=-r}^r e^{-\beta b \sigma \Delta_k},
$$
where $[-r,r]$ is the allowed values of $\Delta_k$.  In the unrestricted case
$r=\infty$, the interfacial free energy per column is
$
{\cal F}=T \ln [ e^{\beta b \sigma}\tanh(b \sigma/(2T))].
$
The zero crossing of ${\cal F}$ gives an estimate of the critical 
temperature.  In the case of the two-dimensional Ising model this zero crossing
gives the exact critical temperature.  This is somewhat fortuitous, but 
nevertheless useful.

\begin{figure}
\centerline{\epsfxsize=3.1in
\epsffile{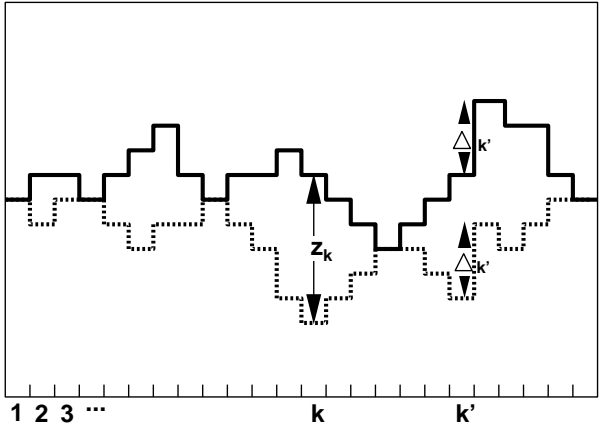}}
\vskip 0.1true cm
\caption{Two solid on solid interfaces.  The interfaces have a negative binding
energy causing them to want to stick but they cannot cross.  This 
``no crossing'' condition results in an entropic repulsion which pushes the 
interfaces apart at high enough temperature.  $z_k$ is the separation of the 
interfaces at the $k$'th step and $\Delta_k$ is the number of steps the 
interfaces take in the same direction at the $k$'th step.}
\label{two_sos}
\end{figure}

In the continuum limit, the problem of two interfaces can usually be 
broken down into a center of mass part and an independent part involving 
the separation of the two interfaces.  We would prefer, however, to work with
a discrete model with parameters input from the energy calculations of the 
appropriate bent domain walls.  We were unable to find the solution to such
a model in the literature, so we present one here.  Questions that we are
interested in are whether or not the two interfaces are bound and whether or
not unbinding occurs before or after the free energy of the walls becomes 
negative.  To answer these questions we consider the following model for two 
SOS walls shown in Figure \ref{two_sos}:
\begin{eqnarray}
{\cal H}_{double} \{\Delta,z\}&=& 
	\sum_k \{(2b\sigma+u_\parallel\delta_{z_k,0})
	+b\sigma |z_k-z_{k-1}|\nonumber\\
& & \qquad	 +(2b\sigma+u_\perp \delta_{z_k,0}) \Delta_k\}.
\label{HdoubleSOS}
\end{eqnarray}
where $z_k$ is the separation of the walls ($z_k \geq 0$), and
$\Delta_k$ is the number of vertical steps the two walls take in the same
direction in the k'th column ($-\infty < \Delta_k <\infty$).  $u_\parallel$
and $u_\perp$ are the binding energies parallel and perpendicular
to the wall.   

The partition function is
\begin{eqnarray}
{\cal Z}&=&\sum_{\{z_k\}} \prod_{k=1}^L e^{-\beta b \sigma |z_k-z_{k-1}|}
e^{-\beta (2 b \sigma+u_\parallel \delta_{z_k,0})} \nonumber\\
& & \,\times\{ (1+|z_k-z_{k-1}|)+\sum_{\Delta_k \neq 0} e^{-\beta(2 b \sigma+u_\perp \delta_{z_k,0})|\Delta_k|}  \}\nonumber\\
& &
\end{eqnarray}
The $(1+|z_k-z_{k-1}|)$ comes from the fact that for $\Delta_k=0$ there are 
$|z_k-z_{k-1}|+1$ ways to divide the change $|z_k-z_{k-1}|$ between the two 
lines.  Summing over $\Delta_k$ leaves the partition function in the form of a 
transfer matrix:
\begin{eqnarray}
{\cal Z}&=&\sum_{\{z_k\}} \prod_{k=1}^L e^{-\beta b \sigma |z_k-z_{k-1}|}
\{\delta_{z_k,0}(|z_k-z_{k-1}|\nonumber\\
& & \qquad\quad+\coth[\beta(b\sigma+u_\perp/2)])e^{-\beta(2b\sigma+u_\parallel)} \nonumber\\
& & \qquad\quad+ (1-\delta_{z_k,0})(|z_k-z_{k-1}|+\coth\beta b\sigma)e^{-2b\sigma}\} \nonumber\\
&=& \sum_{\{z_k\}} \prod_{k=1}^L T_{z_k,z_{k-1}}
\label{part_doub}
\end{eqnarray} 
Unfortunately, we were unable to solve the general case analytically.  However,
restricting $z_k-z_{k-1}$ to $0$ or $\pm 1$, we can derive the eigenvalues and
 eigenvectors of the matrix $\hat {\bf T}$ explicitly.  A ground state 
eigenvector $\psi_\mu(z)$ $=$ $e^{-\mu z}$, where $1/\mu$ is the localization 
length, or typical distance separating the lines, characterizes the bound 
state of the two lines.  $\psi_\mu(z)$ is found by first finding the 
eigenvalue $\lambda_\mu$ (from the defining equation 
$(\hat{\bf T} \psi_\mu)_z=\lambda_\mu \psi_\mu(z)$) for $z>0$. $\mu$ 
is then obtained from the eigenvalue equation for $z=0$.  This gives $e^\mu$
as the solution to the quadratic equation,
\begin{eqnarray}
& &(1+\coth \beta b \sigma)e^{2 \mu}\nonumber\\
& &\,+e^{\beta b \sigma}\left[\coth\beta b \sigma-
e^{-\beta u_\parallel}(1+2 e^{-\beta(2 b \sigma+u_\perp)})\right]e^\mu 
\nonumber\\
& &\,+\left[1+\coth\beta b \sigma-2 e^{-\beta u_\parallel}(1+e^{-\beta(2b\sigma+u_\perp)})\right]=0.
\end{eqnarray}
$\mu=0$ defines the unbinding transition at $T_b$.  The more general case, 
$|z_k-z_{k-1}|<N$ with $N$ a large number (typically about 1000), 
can be easily solved numerically and is not that different from the restricted
case discussed above.  The values quoted in the text are from such a numerical
calculation.

\section{Monte Carlo Simulation of Continuous Spin Systems}
\label{ap:mc}
A reasonable introduction to Monte Carlo techniques is given in 
\cite{BinderHeermann}.  However, some of the implementation techniques 
suggested in this book are out of date and should be taken with a lump of salt.
Most simulations of frustrated spin systems described in the literature appear 
to have used a rather poor updating scheme leading to very long autocorrelation
 times.  We use a heat bath scheme described below which seems to be a couple 
of order of magnitude faster than these standard schemes near the critical 
point.  This is not to say that other heat bath schemes have not been
used, it is just that such works almost never describe any details of
how this is done, a problem we shall try to rectify here. To make efficient 
use of the data generated in a Monte Carlo simulation one should make use of 
the histogram techniques of References \cite{FerrenburgSwend,LeeKosterlitz}. 

\subsection{Sampling}

Formally, the task of statistical mechanics is to compute from the model 
Hamiltonian ${\cal H}$ the desired average properties,
\begin{equation}
\langle A(\{\theta_{ij}\}) \rangle = {1 \over {\cal Z}} \int d\{\theta_{ij}\} A(\{\theta_{ij}\}) \exp \left[ -{\cal H}(\{\theta_{ij}\})/T \right],
\end{equation}
where states are weighted with the normalized Boltzmann distribution
\begin{equation}
p(\{\theta_{ij}\})={1 \over {\cal Z}}\exp \left[ -{\cal H}(\{\theta_{ij}\})/T \right].
\label{boltz}
\end{equation}

While this gives a formally exact description of the probability distribution,
we are not really interested in such detailed information, nor is it possible 
to carry out the integrations in the high-dimensional space required in the 
thermodynamic limit.  The dimension of the space can be reduced somewhat by
making use of finite size scaling to extrapolate from small systems ($L<100$)
to the thermodynamic limit.  Even for these smaller $L$, it is still not 
possible to numerically integrate the system based on any sort of 
discretization scheme.  One instead uses Monte Carlo integration which is 
simply to pick $N$ sets of $\{\theta_{ij}\}$ randomly distributed according to
(\ref{boltz}) and then 
\begin{equation}
\langle A(\{\theta_{ij}\}) \rangle \approx {1 \over N} \sum_{l=1}^N A(\{\theta_{ij}\}_l).
\end{equation}
If the $\{\theta_{ij}\}_l$ are independent and $A(\{\theta_{ij}\})$ is 
distributed in a Gaussian distribution with variance $\sigma^2$ then the error 
in $\langle A \rangle$ calculated in this manner is $\sigma/N^{1/2}$.

In practice, the knowledge of how to pick independent random numbers 
distributed according to (\ref{boltz}) is quite close to knowing how to 
solve the problem exactly.  In general, we must give up on the idea of
independent random numbers and instead construct a Markov process where
each state $\{\theta_{ij}\}_{l+1}$ is constructed from a previous state 
$\{\theta_{ij}\}_l$ via a suitable transition probability
$W(\{\theta_{ij}\}_l \longrightarrow \{\theta_{ij}\}_{l+1})$.  A sufficient
condition for the distribution function $P(\{\theta_{ij}\})$ of states 
generated to converge to (\ref{boltz}) in the limit $N\rightarrow \infty$, is 
for the transition probability to satisfy detailed balance:
\begin{equation}
{W(\{\theta_{ij}\}_l \longrightarrow \{\theta_{ij}\}_{l'}) \over 
W(\{\theta_{ij}\}_{l'} \longrightarrow \{\theta_{ij}\}_l)} = 
\exp \left(- {\delta {\cal H} \over T}\right),
\label{balance}
\end{equation} 
where 
$\delta {\cal H}={\cal H}(\{\theta_{ij}\}_{l'})-{\cal H}(\{\theta_{ij}\}_l)$.
Note that equation (\ref{balance}) must be satisfied for {\it all} possible
moves $l\rightarrow l'$ in order to be ergodic.

This still leaves many choices for the move.  Ideally, one would like to 
change many degree's of freedom simultaneously, unfortunately in the absence 
of any cluster routines for frustrated systems, one is left with single site
updating moves. (Alternatively one can simulate a Langevin equation to change
all degree's of freedom simultaneously, but by a small amount.  Even Langevin
dynamics are not unique, and the dynamics which are supposed to be appropriate
for superconducting arrays \cite{Faloetal} was found to have longer 
autocorrelation times than the Monte Carlo method we ended up using.)  One 
particularly poor, but popular, method of updating continuous degrees of 
freedom involves picking a new $\theta_{ij}$ completely at random, or in an
interval about it's previous value, and then accepting or rejecting the move
based on whether another random number is above or below 
$\exp \left(- {\delta {\cal H} \over T}\right)$.  This can give extremely long
autocorrelation times, and leads to a high number of rejected moves in the low
temperature state.  One would have to apply this same step numerous times to
the same spin just to equilibrate it with it's nearest neighbors.

An ideal single site updating step would pick $\theta_{ij}$ according to the
conditional Boltzmann probability $p(\theta_{ij})$ for $\theta_{ij}$ given the 
knowledge of the neighboring spins $\{\theta_{i,j\pm 1},\theta_{i\pm 1,j}\}$.  
For our frustrated $XY$ model this is
\begin{eqnarray}
p(\theta_{ij})&=&{1 \over C} \exp\left[
(\cos(\theta_{i,j+1}-\theta_{ij}+A_{ij}^{i,j+1}) \right. \nonumber\\
& &\quad\qquad +\cos(\theta_{i,j-1}-\theta_{ij}+A_{ij}^{i,j-1}) \nonumber\\
& &\quad\qquad +\cos(\theta_{ij}-\theta_{i+1,j}+A_{i+1,j}^{ij})\nonumber\\
& &\quad\qquad \left.+\cos(\theta_{ij}-\theta_{i-1,j}+A_{i-1,j}^{ij}))/T\right]\nonumber\\
 &=& {1 \over I_0 \left( {h\over T}\right)} \exp\left( {h\over T} \cos (\theta_{ij}-\delta) \right),
\label{condprob}
\end{eqnarray}
where
\begin{eqnarray}
h &=& \sqrt{x^2+y^2},\nonumber\\
\delta &=& \arctan(x/y),\nonumber\\
x &=& \sin(\theta_{i,j+1}+A_{ij}^{i,j+1}) 
+\sin(\theta_{i,j-1}+A_{ij}^{i,j-1}) \nonumber\\
& & +\sin(\theta_{i+1,j}-A_{i+1,j}^{ij})+
\sin(\theta_{i-1,j}-A_{i-1,j}^{ij}),\nonumber\\
y &=& \cos(\theta_{i,j+1}+A_{ij}^{i,j+1}) 
+\cos(\theta_{i,j-1}+A_{ij}^{i,j-1}) \nonumber\\
& & +\cos(\theta_{i+1,j}-A_{i+1,j}^{ij})+
\cos(\theta_{i-1,j}-A_{i-1,j}^{ij}),
\end{eqnarray}
and $I_0(x)$ is the zeroth order modified Bessel function.

\begin{figure}
\centerline{\epsfxsize=3in
\epsffile{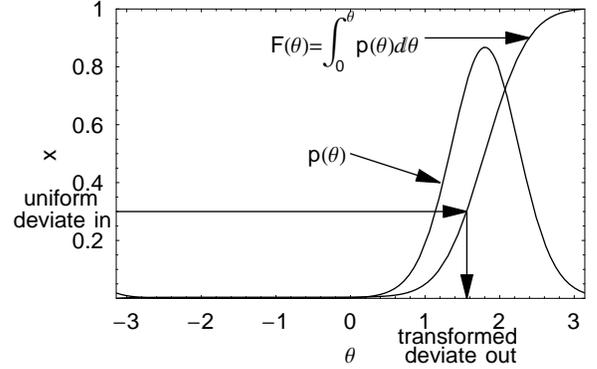}}
\vskip 0.1true cm
\caption{Transformation method for converting a uniform deviate $x$ into a
random deviate $\theta$ distributed according to the function $p(\theta)$.}
\label{trans}
\end{figure}

An excellent reference for the next step can be found in \cite{Devroye}.
In order to generate a distribution of $\theta$ with $p(\theta)$ given by 
(\ref{condprob}), one first generates a uniform deviate $x$ (independent
uniformly distributed random number between 0 and 1) and makes use of the 
fundamental transformation law of probabilities, which simply tells us
\begin{equation}
|p(\theta) d\theta|=|dx|.
\end{equation}
So we need to solve
\begin{equation}
{dx \over d\theta}=p(\theta).
\end{equation}
The solution of this is $x=F(\theta)$, where $F(\theta)$ is the indefinite
integral of $p(\theta)$.  The desired transformation which takes a uniform
deviate into one distributed as $p(\theta)$ is therefore
\begin{equation}
\theta(x)=F^{-1}(x)
\end{equation}
where $F^{-1}$ is the inverse function to $F$.  This process is illustrated
in Figure \ref{trans}.

Unfortunately, $F$ (and $F^{-1}$) can only be computed numerically.  In order
to implement the method we used look-up tables and interpolation.  On
systems where integer operations are much faster than floating point 
operations, things can be speeded up considerably by discretizing the 
$\theta_{ij}$ (for instance one can take the integers $0$ to $524288$
to correspond to $0$ to $2 \pi$) and then storing all possible values
of the sinusoidal functions that can occur (all $524288$ values).
This requires some storage capacity (about 64 Mbyte for our
implemention) but this should not be onerous for any machine that one
would consider doing such simulations on.  One should note that some
machines can compute trigonometric functions in only a few clock
cycles and therefore it may be faster than a look-up call to memory. The 
resulting code took about twice as long per Monte Carlo step (MCS) to run as 
the simple ``pick at random and then reject'' method, but this loss is more 
than compensated for by the orders of magnitude improvement in correlation 
times.  There is still considerable freedom in the order in which subsequent
lattice sites are selected.  Naively, one would think that, as long as all
sites are visited on some pseudo regular basis, that the order is unimportant.
While this is true in the sense that the order is unimportant for eventually
reaching equilibrium, the order can have a huge impact on how fast you get
there.  The slowest (in the sense of long correlation times) method is to
select sites at random.  One can significantly reduce (by a factor of up to 
about $L$ depending on temperature) correlation times by going through the 
lattice in typewriter fashion or a mixture of random and typewriter ordering.  However, one must go through in different 
directions (alternate left-right-up-down with up-down-left-right etc.) in 
order for the correlation times to be isotropic (i.e. have the same 
correlation time for say $Y$ measured in both the $x$ and $y$ direction).
To ensure the accuracy of the implementation, the code was tested against 
published results for the $f=0$ and $f=1/2$ cases.

\subsection{Error Analysis}

Suppose we make $N$ successive observations $A_\mu, \mu=1,\cdots,N$, of a
quantity $A$ in our simulation.  If the distribution of the fluctuations in
$A$ is Gaussian (this is {\it not} true for all the parameters measured), then
the expectation value of the square of the statistical error, which in this 
case is the variance, is
\begin{eqnarray}
\langle (\delta A)^2\rangle&=&\left\langle \left[{1\over N} \sum_{\mu=1}^N (A_\mu-\langle A \rangle) \right]^2 \right\rangle\nonumber\\
&=& {1\over N^2}\sum_{\mu=1}^N\langle(A_\mu-\langle A\rangle)^2\rangle\nonumber\\
& &\quad+{2\over N^2}\sum_{\mu_1=1}^N \sum_{\mu_2=\mu_1+1}^N (\langle A_{\mu_1}A_{\mu_2}\rangle -\langle A\rangle^2) \nonumber\\
%&=&{1\over N}\left[\langle A^2\rangle-\langle A \rangle^2\right.\nonumber\\
%& &\left.\quad+2\sum_{\mu=1}^N\left(1-{\mu\over N}\right)(\langle A_0 A_\mu\ra%ngle -\langle A \rangle^2)\right]\nonumber\\
&=& {1\over N}(\langle A^2\rangle-\langle A \rangle^2)\nonumber\\
& & \quad\times\left[1+2\sum_{\mu=1}^N\left(1-{\mu\over N}\right){\langle A_0 A_\mu\rangle -\langle A \rangle^2\over \langle A^2\rangle-\langle A \rangle^2}\right].\nonumber\\
\label{auto_der}
\end{eqnarray}
The autocorrelation function for $A$ is defined as
\begin{equation}
\phi_A(t_\mu)={\langle A_0 A_\mu\rangle -\langle A \rangle^2\over \langle A^2\rangle-\langle A \rangle^2}.
\end{equation}
where we associate the time $t_\mu$ with step $\mu$.  Note that $\phi_A(0)=1$ 
and $\phi_A(t)$ decays to zero as $t\rightarrow \infty$.  The autocorrelation
time $\tau_A$ is defined as
\begin{equation}
\tau_A=\sum_{\mu=1}^\infty \phi_A(t_\mu).
\end{equation}  
For an exponential relaxation, $\tau_A$ is the relaxation time, 
so that for times $t_\mu \gg \tau_A$, $\phi_A(t_\mu)$ is very small.  If 
$N \gg \tau_A$ we can, therefore neglect the term involving $\mu/N$ in 
Eq.(\ref{auto_der}) and one obtains
\begin{equation}
\langle (\delta A)^2\rangle={1\over N}(\langle A^2\rangle-\langle A \rangle^2)(1+2 \tau_A).
\label{var}
\end{equation}
Thus, our $N$ correlated measurements are equivalent to $N/(1+2 \tau_A)$ 
independent measurements, something that must be taken into account when
calculating errors.

The concept of self-averaging (or lack of) is extremely important in correctly
estimating errors from Monte Carlo simulations with disorder.  Suppose we 
measure $A$ and calculate it's statistical error using 
$\sqrt{\langle (\delta A)^2\rangle}$ from Eq.(\ref{var}).  If 
$\sqrt{\langle (\delta A)^2\rangle}$ reduces to zero if $L\rightarrow \infty$ 
(and $N/(1+2 \tau_A)$ fixed) we say $A$ exhibits self-averaging.  If, on the 
other hand, $\sqrt{\langle (\delta A)^2\rangle}$ reaches an $L$-independent 
nonzero limit, we say $A$ exhibits a lack of self-averaging.   
Random systems exhibit a lack of self averaging near the critical 
point \cite{selfaverage}.  In fact, the distribution of most quantities 
(over realizations of disorder) is not even Gaussian, making the use of 
$\sqrt{\langle (\delta A)^2\rangle}$ as a measure of the statistical error 
somewhat questionable.

In calculating errors we make use of, among other things, the bootstrap 
resampling technique described in \cite{bootstrap} and more compactly in
\cite{NumRec}.  From the set of data ${\cal D}_0$ produced by our Monte Carlo 
simulation we calculate a set ${\bf x}_0$ of parameters such as the 
energy, order parameter, etc.  Due to the random sampling, ${\cal D}_0$ is not
a unique realization of the true parameters ${\bf x}_{true}$.  With different
initial conditions or other slight variations we could have measured any of
an infinite number of other realizations ${\cal D}_1$, ${\cal D}_2$, $\cdots$.
Although the set ${\bf x}_0$  is not the true one ${\bf x}_{true}$, we assume 
that the shape of the probability distribution ${\bf x}_i-{\bf x}_0$,
 is the same, or very nearly the same, as the shape of the
probability distribution ${\bf x}_i-{\bf x}_{true}$.  This is not an assumption
that ${\bf x}_0$ and ${\bf x}_{true}$ are the same, it is just assuming that
the way in which random errors enter the simulation does not vary rapidly as a 
function of ${\bf x}_{true}$, so that ${\bf x}_0$ can serve as a reasonable
surrogate.  

Suppose we have in some way obtained a set of equivalent realizations of our
data. For each realization ${\cal D}_j$ we calculate the parameters 
${\bf x}_j$ in the same way as we obtained ${\bf x}_0$ from ${\cal D}_0$. 
Each simulated measured parameter set yields a point ${\bf x}_j-{\bf x}_0$.
If we simulate enough data sets we can map out the desired probability 
distribution for the parameter space.  As mentioned above, this distribution
of parameters is not necessarily Gaussian so we require some means of 
defining what we mean by the statistical error.  We take the statistical error
to be width of the confidence region that contains 68\% of the data (i.e. the
confidence region is defined by the interval $x_0\pm\sigma$ where, given
the set of realizations of the parameter $x$, 68\% of the $x_j$ lie between
$x_0-\sigma$ and $x_0+\sigma$).  In this way, if our distribution is Gaussian,
our definition of the error is just the standard deviation, as one would want
for compatibility with the standard case.

It only remains to explain how we obtain ``a set of equivalent realizations of 
our data''.  The bootstrap method \cite{bootstrap} used the actual data set 
${\cal D}_0$, with it's $n=N/(1+2 \tau_A)$ ``independent'' data points, to
generate any number of synthetic data sets ${\cal D}_j^S$, with $n$ data 
points.  The procedure is simply to draw $n$ data points at a time {\it with
replacement} from the set ${\cal D}_0$.  For the bond disordered systems this
includes bootstrap resampling of the set of realizations of bond disorder, as
well as bootstrap resampling of the data from an individual realization of 
disorder.  The basic idea behind the bootstrap is that the actual data set,
viewed as a probability distribution, is the best available estimator of the 
underlying probability distribution.

\begin{table}
\narrowtext
\begin{tabular}{ccc}
{\it domain wall type} & \multicolumn{2}{c} {\it energy per unit length} \\  
	       & $f=1/3$        & $f=2/5$        \\ 
\hline
herringbone-0  & $0.05673742$ J & $0.08611726$ J \\
herringbone-1  & $0.19503538$ J &       -        \\
shift-by-1     & $0.11419998$ J & $0.15889929$ J \\
shift-by-2     & $0.16666667$ J & $0.16612232$ J \\
shift-by-3     &      -         & $0.14764859$ J \\
\end{tabular} 
\caption{Domain wall energies for stable domain wall structures (i.e. walls 
which produce a vortex pattern consistent with 
$\delta {\cal H}/\delta \theta_j$ $=$ $0$ for every $\theta_j$).  The $n$ in 
herringbone-$n$ denotes the associated shift where $n$ $=$ $0$ is the standard 
herringbone.}
\label{wall_energies}
\end{table}

\end{multicols}
\end{document}